\documentclass[12pt]{iopart}
\usepackage{graphicx}
\usepackage{todonotes}
\usepackage{hyperref}
\usepackage{epigraph}
\usepackage{comment}
\RequirePackage{xcolor}
\definecolor{green}{rgb}{0,.9,.15}
\definecolor{violet}{rgb}{.8,.12,.86}

\begin{document}
\raggedright
\title[Assessing the Performance of Two Analytic Gravitational Waves Models]{Merging Black Holes: Assessing the Performance of Two Analytic Gravitational Waves Models}
\author{Dillon Buskirk \footnote{Present address: Department of Physics \& Astronomy, University of Kentucky, Lexington, KY 40506, USA} and Maria C. Babiuc Hamilton \footnote{Corresponding author: \href{mailto:babiuc@marshall.edu}{babiuc@marshall.edu}}}
\address{Department of Physics, Marshall University, Huntington, WV 25755, USA}
\vspace{10pt}
\begin{indented}
\item[]
\end{indented}
\epigraph{I would rather have questions that can't be answered than answers that can't be questioned.}{Richard P. Feynman}
\begin{abstract}
Merging black holes produce the loudest signal in the detectors. 
However, this is the most difficult signal to accurately predict with analytical techniques. 
Only computer simulations can account for the nonlinear physics during the collision, but they are inherently complex, costly, and affected by numerical errors.
In order to bypass this problem, two analytical models for the merger have been developed: the Implicit Rotating Source (IRS) and the newer Backwards one Body (BoB).
In this work, we assess the performance of the BoB model by comparing it with the older IRS model and with the numerical data, identifying its strengths and weaknesses. 
Our main finding reveals discrepancies in amplitude, but overall excellent accord in frequency. 
The BoB model is comparable with the IRS and NR simulations, having the added advantage that it depends only indirectly on numerical data, it accounts for spin, and it offers a seamless fit with the analytical formalisms for the inspiral. 
By independently evaluating and testing those models, we bring evidence of their reproducibility, thus upholding high scientific standards, and make available our implementation, enabling readers to evaluate our results themselves.
\end{abstract}
%
%
%
%
%

\section{Introduction}
\label{sec:intro}
When two black holes collide, they release an enormous amount of energy in gravitational waves (GW), qualifying such an event as one of the most energetic phenomena in our universe. 
Indeed, it was estimated that the GW signal detected on September 14, 2015 \cite{arXiv:1602.03837, arXiv:1608.01940} carried away the energy equivalent of three solar masses, in a fraction of a second. This event is more powerful than the combined luminosity of all the stars and galaxies in the observable universe.
This discovery, detected by the Laser Interferometer Gravitational-Wave Observatory (LIGO) \cite{arXiv:1411.4547},  created a new branch of science, named Gravitational-wave Astronomy.
From 2015 to 2021, over 90 GW detections coming from mostly binary black holes (BBH) collisions were reported by the LIGO and Virgo observatories \cite{arXiv:2111.03606, arXiv:2112.06878}. 

Although tremendously energetic near the source, by the time they reach Earth, GW are so weak that they perturb the LIGO detector arms by only $10^{-18} \texttt{m}$. 
Detecting such a small signal is extremely challenging and depends on the accuracy of the GW models used to filter it from the inherent noise that affects the detector.
The detection of GW is important because it reveals a population of binary black holes (BBH) not visible in the electromagnetic spectrum, which in turn constrains and drives the development of accurate physical models predicting the formation and evolution of binary systems
 \cite{arXiv:2010.14533, arXiv:1811.12940}.
But we cannot tease this information from the collected data unless we have accurate templates to filter the signal from the noise.

The framework needed to provide GW templates is Einstein’s theory of General Relativity (GR). 
In this theory, time can no longer be considered invariant, and a complete analytical solution of the gravitational system under consideration requires knowledge of the entire four dimensional spacetime.
This demand makes the problem particularly challenging to solve analytically, and solutions to Einstein’s equations cannot be found except in highly symmetric cases \cite{book:exactgr}.
As expected then, for spacetimes as complex as those of BBH systems, we must rewrite Einstein's equations of GR in forms suitable for numerical integration and then resort to computer simulations to obtain accurate results (see \cite{arXiv:1411.3997} for a review). 

The next generation of GW detectors network, which includes both ground-based observatories such as Cosmic Explorer \cite{arXiv:2012.03608}, or Einstein Telescope \cite{arXiv:1912.02622}, and space-borne instruments such as LISA \cite{arXiv:1907.11305, arXiv:2001.09793}, DECIGO \cite{arXiv:1802.06977}, or TianQin \cite{arXiv:2008.10332}, are in various stages of development. 
When those observatories come online, they will detect GW at a much faster pace and enhanced sensitivity, reaching so deep into space as to reveal mergers at redshifts of 10 and above, thus shedding light on black holes formed by either the first stellar population of stars in the Universe, or even on primordial black holes, formed by quantum fluctuations shortly after the Big Bang \cite{arXiv:1907.04833}. 
It is estimated that those future instruments will report binary detections at thousands of events per year \cite{arXiv:2109.09882}, greatly increasing the demand in template generation.

In order to bypass the complexity and the cost of numerical simulations, analytical models have been developed for creating complete GW templates, from the time the signal enters the detector's band, continuing with the binary black hole (BBH) collision and finishing with the ringdown.
In those approximations, increasingly higher-order expansions are added to Newton’s law of gravity in terms of the orbital velocity, to work towards solutions of Einstein’s equations \cite{arXiv:1607.05661}, or high-order polynomials are fitted directly to the existent numerical relativity  (NR) data \cite{arXiv:1611.00332}. 
More recently, efforts are made to compute GW templates using quantum scattering amplitudes \cite{arXiv:2107.10193}.
They are also very useful, because they allow us to employ approximate functions that are much simpler to work with, but contain the physical information we need to a high degree of accuracy, to quickly generate GW templates. 
Indeed, analytical models are easy to generate, are not plagued by numerical errors inherent to computer simulations \cite{arXiv:1808.06011} and do not rely on interpolation, built-in the NR-based surrogate GW models \cite{arXiv:2101.11608}. 

Even so, analytical models do present the disadvantage that they are lengthy, cumbersome to build, and when it comes to the merger phase, they rely entirely on NR results to encompass the nonlinear physics of the collision, which becomes thus the least accurate part of those models, although it is the loudest signal in the detector.
There are two independent analytical approaches that are concerned with modeling only the merger, (1) the \emph{implicit rotating source} (IRS) model \cite{arXiv:0805.1428, arXiv:1107.1181, arXiv:1212.0837, arXiv:1403.0561, arXiv:1403.7754}, and (2) the recently developed \emph{backwards-one-body} (BoB) model \cite{arXiv:1810.00040, arXiv:2001.11412}. 
They rely on the remarkable simplicity and continuity of the GW around the merger, which is surprising in view of the violence of the collision and of the nonlinear nature of the GR interactions. 
Among those two, the BoB model holds great potential for GW data analysis.    
In this work we bring under scrutiny this analytical model and ask the questions: How does it compare with the IRS model in terms of implementation and results? How does it uphold its predictions against the numerical GW templates, known to correctly model the merger, because they include all the necessary physics?

Our objective is to identify the strengths and weaknesses of the BoB analytical model employed for the merger, and to find out if it approximates correctly the GW signal, by comparing it to the older IRS model and to the GW calculated by NR.
To achieve this purpose, we must first assemble the essential ingredients, namely the mass of the final black hole, spin and quality factor, which are required as initial data by the BoB model.
We dive into an independent comparison of the methods available to approximate the spin, mass and angular frequency of the final black hole, because this is instrumental in assuring the accuracy of the merger models.
We proceed next by independently implementing the IRS and BoB approaches using the \texttt{Wolfram Mathematica} software, then we thoroughly compare them, and continue with weighing the analytical models against NR results. 

Our main finding is that the performance of the BoB model is comparable with the IRS and NR simulations, having the added advantage that it depends on less numerical parameters, it includes the spin, and it offers a seamless continuation with the analytical formalisms for the inspiral into the merger and ringdown.
By independently evaluating and testing those models, we not only bring evidence of their reproducibility, an essential requirement for the upholding of high scientific standards, but with our findings, we enable readers to discern their similarities and differences, and to evaluate our results themselves.

We start our presentation with the theoretical framework in \Sref{sec:theory}, by first describing the intricacies of obtaining the initial data necessary to build the analytical merger models. 
We follow in Sub\sref{ssec:BoB-IRS} with a short explanation on how NR calculates the strain of the GW,  then we summarize the ansatz entering in the two merger models: Sub\sref{sssec:IRS} describes IRS, and Sub\sref{sssec:BoB} reviews BoB, to arrive at the strain of the GW emitted during merger.
Once the theoretical  framework is laid out clearly, it is time for us to present our results in \Sref{sec:results}. 
We start by detailing our calculations to obtain the final spin, mass and orbital ringdown frequency of the final black hole in Sub\sref{ssec:before}.
We continue in Sub\sref{ssec:BoBcomp} by implementing the two analytical methods describing the strong-field region of the merger, and 
we compare them both in frequency, phase and amplitude with the NR data provided by the Simulating eXtreme Spacetimes project (SXS) \cite{arXiv:1304.6077, arXiv:1904.04831}, then we compare their strain for three mass ratios of non-spinning black holes.
Finally, we asses in Sub\sref{ssec:BoBspin} the performance of the BoB model at high spins against the corresponding NR data from the SXS catalogue. 
We implement our work in a set of new, freely available \texttt{Wolfram Mathematica} notebooks.
We conclude in \Sref{sec:conclude} with a summary of our findings and suggest extensions of our work. 

Note that we will use geometrized units where $G = c = 1$ throughout, which allows us to measure time, space and mass in terms of the mass of the binary system at infinity, defined as: $M = m_1 + m_2$, where $m_1$ and $m_2$ are the masses of the individual black holes.
In order to recover the required physical units, we simply multiply mass with $M_\odot (\texttt{kg})= 1.98892 \times 10^{30} \texttt{kg}$, space with $ M_\odot (\texttt{km}) =  1.477~\texttt{km}$, and time with $M_\odot (\texttt{s}) = 4.92674\times 10^{-6} \texttt{s}$. 


\section{Theoretical Framework}
\label{sec:theory}

\subsection{Initial Considerations}
\label{ssec:initial}
Before we introduce the models for the merger, we must start with the end in mind, because we require first the mass, spin and ringdown frequency of the final black hole. 

\subsubsection{Final Spin}
\label{sssec:spin}
Electromagnetic X-ray observations of black holes show that they are spinning very fast, with most of their dimensionless spin in the $(0.5 - 0.8)$ range \cite{arXiv:1903.11704}. The dimensionless spin is defined as $\chi = S/m^2$ where $S$ is the spin angular momentum and $m$ is the black hole mass. 
Although the observed X-ray highly-spinning binaries might not merge within the Hubble time \cite{arXiv:2207.14290}, the current GW data supports a population of BBH with non-negligible spin magnitudes \cite{arXiv:2205.08574, arXiv:2205.12329}.
There are various proposed mechanisms through which BBH systems that are close enough to merge within the Hubble time can acquire high spins, either through the isolated formation channel \cite{arXiv:2010.00078}, or through hierarchical mergers in dense stellar clusters  \cite{arXiv:2105.03439}.
If each individual BH in the binary does spin, the interactions between them can make the orbital angular momentum precess and nutate, which gets imprinted in the GW signal \cite{arXiv:gr-qc/9402014, arXiv:gr-qc/9506022, arXiv:2103.03894}. 
The spin affects the orbital dynamics of the merger and interacts with the spacetime through relativistic frame-dragging, a well-known and tested \emph{gravitomagnetic} GR effect \cite{arXiv:1403.7377, arXiv:1910.09908}.

We should not be surprised to find out that, due to the conservation of the angular momentum, even when the two bodies do not spin, the final black hole acquires an appreciable spin. This spin is usually aligned with the orbital angular momentum at large separations \cite{arXiv:0904.2577}, especially if the binary coevolved, because interactions such as spin-orbit and spin-spin coupling will tend to align the individual spins with each other and with the orbital angular momentum \cite{arXiv:1002.2643, arXiv:1312.5775, arXiv:1406.7295, arXiv:1609.05933}.
Thus, although the misalignment between the orbital angular momentum and the individual spins could lead to interesting effects such as orbital precession and recoil \cite{arXiv:1503.07536, arXiv:1610.09713}, 
we will take into consideration in this work only the components of the individual spins parallel to the orbital momentum, leaving the problem of their general orientation to a future study.
This is not as restrictive as it seems and it can be applied also to spins with arbitrary orientation, because the misalignment between the orbital angular momentum and the individual spins has only a small impact on the magnitude of the final black hole's spin \cite{arXiv:1612.02340, arXiv:1809.01401}.  
The loss of accuracy introduced by this simplification is insignificant for the typical Advanced LIGO (ALIGO) and Virgo observations \cite{arXiv:1908.00555, arXiv:1909.05466, arXiv:2010.04131}, even for unusual mergers \cite{arXiv:1602.03840, arXiv:2009.01190, arXiv:2009.01066, arXiv:2009.05461, arXiv:2010.12558}.

We denote the dimensionless individual spins of the black holes entering the merger with $\chi_i= S_i/m_i^2$, where $i = 1, 2$.
Here, $S_i$ is the spin angular momentum and $m_i$ represents the mass of each black hole.
Those spins are usually grouped into an \emph{effective spin} defined as a mass-weighted combination of the individual spins:
\begin{equation}
\chi_{eff}= \frac{m_1^2 \chi_1+m_2^2 \chi_2}{m_1^2+m_2^2}.
\label{eq:chieff}
\end{equation}
This combination is not unique. 
Our choice corresponds to fits of the post-merger BH spin to NR simulations \cite{arXiv:1506.03492, arXiv:0710.3345, arXiv:0712.3541, arXiv:1508.07250, arXiv:1605.01938}, which differs from the effective inspiral spin parameter commonly used in the literature \cite{arXiv:2103.03894, arXiv:gr-qc/0103018, arXiv:0803.1820}.

The spin of the final black hole $\chi_{f}$ will depend both on the masses of the individual black holes and on the magnitudes of their spins. 
Although this physical dependence is nonlinear and can only be determined from numerical simulations, it can subsequently be modeled analytically with the help of a fitting polynomial to the numerical results.  
We will use in this work the simple expression for the final spin given in \cite{arXiv:0904.2577}, written in function of $\chi_{eff}$ and the dimensionless quantity $\eta = \mu/M$ called \emph{symmetric mass ratio}, where $\mu =m_1 m_2/M$ is the \emph{reduced mass}:
\begin{equation}
\chi_{f} =\sum^3_{i,j=0} s_{ij} \eta^i\chi^j_{eff}.
\label{eq:chifin}
\end{equation}
The coefficients $s_{ij}$ are determined by tuning the analytical expression to NR results and by calibrating them to the analytical calculations available for  the extreme mass-ratio regime. 
This formula gives the correct result of $\chi_f = 0.686$ for equal-mass non-spinning binaries and can be applied as well to systems with unequal masses and spins. 
Details of the various coefficients proposed in the expression given by eq.\eref{eq:chifin} can be found in \cite{arXiv:0710.3345, arXiv:0712.3541, arXiv:1605.01938, arXiv:0706.3732}.
We follow \cite{arXiv:1605.01938} to calculate the final spin for equal-mass binaries, and for unequal-mass binaries we implement the formulas given in \cite{arXiv:1406.7295, arXiv:1610.09713} and in \cite{arXiv:1611.00332}, as explained in \ref{appendixA}.

\subsubsection{Final Mass}
\label{sssec:mass}
In GR mass is not an invariant concept and in order to define it independently of coordinates, we must either find a fixed point, or take into account the entire mass of the universe.
Because neither of those alternatives are possible, we must rely on two plausible assumptions, namely that the universe is finite and that objects occupy certain regions in space, in order to define mass. 
This allows us to introduce the notion of a \emph{local mass} for isolated objects, called the \emph{ADM mass} \cite{book:Wald}, which measures the energy of the spacetime, it is independent of the choice of coordinates and coincides with the mass at infinity $M$, when the space becomes \emph{asymptotically} flat.
The problem is even more complicated for a binary system. 
First, some of its mass is used up in the gravitational binding energy that keeps the binary together, thus  the ADM mass of the system will be  smaller than the sum of the individual black holes masses. 
Moreover, we expect mass loss in energy as the system emits GW, causing the ADM mass to decrease even further during the evolution of the binary. 
As result, the final mass of the remnant black hole $M_f$ is smaller than the mass of the system at infinity $M$. 
This final mass can be calculated with the Christodoulou formula once the final spin $S_f$ is known \cite{christodoulou, smarr}.
\begin{equation}
M_f= \sqrt{M_{irr}^2+\frac{S_f^2}{4 M_{irr}^2}} ,~\textrm{where}~M_{irr} = \sqrt{\frac{A_{AH}}{16 \pi}}.
\label{eq:MChris}
\end{equation}
Here, $M_{irr}$ is called the \emph{irreducible} or \emph{Christodoulou mass}, and requires the knowledge of the remnant black hole surface area $A_{AH}$, which can only be deduced numerically, from the properties of the common apparent horizon \cite{arXiv:gr-qc/0206008, arXiv:gr-qc/0306056}.
We are interested in analytic expressions and will take another route, namely we will find estimates for the mass loss through gravitational radiation, then we will subtract it from the ADM mass to obtain the mass of the final black hole.

Intuitively, the strategy followed to calculate the energy carried away by the GW is to fit a second-order Taylor expansion in terms of the mass ratio and the spins of the two black holes to this energy \cite{arXiv:0805.1428, arXiv:0710.3345, arXiv:0706.3732, echeverria, arXiv:gr-qc/0703053, arXiv:0907.0462}:
\begin{equation}
{\tilde E_{GW}} = \sum_{i,j=0}^2 p_{ij}   \eta^i (\chi_1 + \chi_2)^j,
\label{eq:EGW}
\end{equation}
where ${\tilde E_{GW}} = E_{GW}/M_{ADM} $ and is dimensionless.
As expected, the coefficients $p_{ij}$ are obtained by fitting with NR results. 
For equal-mass binaries, this formula predicts that about $3\%$ of the total ADM mass is released as GW energy, with the highest efficiency of about $10\%$ when the individual spins are aligned. 
Our analytical implementation of the remnant black hole mass is given in \ref{appendixB}.
We used four expressions, as in \cite{arXiv:0907.0462, arXiv:1312.5775, arXiv:1406.7295, arXiv:1610.09713}, as well as a fifth one, following the hierarchical approach given in \cite{arXiv:1611.00332}.

\subsubsection{Final Frequency Modes}
\label{sssec:qnm}
We should not be surprised to find out that to correctly estimate the GW frequency during the merger, we must know what are the last notes reverberating by the final black hole as it settles down after the collision.
These are its resonance perturbations, similar to the strike tones of bells when they are hit.
The frequency of those perturbations become complex due to the emission of GW:
\begin{equation}
\omega= \omega_R + \mathrm{i} \omega_I,
\label{eq:omfin}
\end{equation}
with the real part representing the actual frequency of the GW modes $\omega_R=\omega_{lmn}$, while the imaginary part encodes the damping time $\tau_{lmn} = 1/\omega_I$ \cite{arXiv:gr-qc/9909058}.
This kind of perturbation is called \emph{quasi-normal} ringing and is modeled as a linear superposition of exponentially damped \emph{quasinormal modes} (QNM). 
When the final black hole has angular momentum, those natural modes are described by \emph{spin-weighted spheroidal harmonics} ${}_{s} S_{lmn}$ \cite{arXiv:1404.3197}, that reduce to spin-weighted spherical harmonics  ${}_{s} Y_{lmn}$ for non-rotating (Schwarzschild) black holes \cite{arXiv:0905.2975, arXiv:1408.1860}. 
The strain of the GW is written as \cite{arXiv:gr-qc/0512160, arXiv:0905.2975}: 
\begin{equation}
h=\frac{M}{r}\sum_{lmn} {\cal A}_{lmn} e^{\mathrm{i} (\omega_{lmn} t+ \phi)} e^{- t / \tau_{lmn}} {}_{s}{\cal S}_{lmn},
\end{equation}
where $s=-2$ is the spin, $l$ is the principal and $|m| \le l $ the azimuthal index of the multipolar order, while $n$ is the overtone of the mode. 
The $l=m$ modes are more energetic \cite{arXiv:gr-qc/0703053}, with most of the GW's energy ($\approx 95\%$) carried away by the $l =  m = 2$ mode \cite{arXiv:gr-qc/0512160}, while the higher harmonics are much quieter.
For comparison, the next energetic mode ($l = m = 4$) is $\sim 10^{-4}$ smaller than the dominant mode \cite{arXiv:0907.0462}.
{
We will denote the frequency of the orbital mode ($l=2, m=1, n=0$)} by $\Omega_{QNM}$. 
The dominant ($l=2, m=2, n=0$) mode can be derived from this mode, with the simple relations $\omega_{lm} =  m \Omega_{QNM}$. 
Intuitively, $\Omega_{QNM}$ is equal to the orbital frequency of the photons trapped at the light ring (LR) or photon-sphere, moving around a Schwarzschild black hole, or with the orbital frequency of the prograde photon ring in the case of rotating black holes. Thus, the QNM ringing is interpreted as the frequency of the GW trapped in an unstable orbit at the light ring, which is slowly diffusing out in a time measured by the damping time \cite{arXiv:0905.2975}. 
The LR marks the peak of the effective potential and constitutes a barrier that filters out anything inside this region, thus an outside observer detects only the QNM. 

It is known that the damping time is linked with this QNM frequency by the quality factor $Q_{QNM}$ given below, a dimensionless quantity measuring the number of oscillations observed before the mode dissipates \cite{arXiv:0903.0338}.
\begin{equation}
Q_{QNM} = \frac{1}{2} \Omega_{QNM} \tau_{QNM}.
\label{eq:QtauQNM}
\end{equation}
The QNM frequency and quality factor are subsequently completely determined by the mass and spin of the final black hole, and can be calculated precisely, either with the perturbation theory \cite{arXiv:gr-qc/9909058}, or by fitting simple analytical functions to NR data \cite{arXiv:2001.10914}.
An important result in this regard was provided by \cite{mashhoon} where the QNM frequencies are calculated analytically using a variational formalism, applicable to slowly rotating black holes.
A comprehensive analysis of the black hole ringdown oscillation modes is given in \cite{BertiWeb}.
We choose a simple analytical fit for the pair ($\Omega_{QNM}, Q_{QNM}$) constructed in \cite{arXiv:gr-qc/0512160, arXiv:0905.2975}:
\begin{equation}
\label{eq:MfOQNM}
M_f \Omega_{QNM} = f_1 + f_2(1 - \chi_{f})^{f_3},
\end{equation}
\begin{equation}
\label{eq:QQNM}
Q_{QNM} = q_1 + q_2(1 - \chi_{f})^{q_3}, 
\end{equation}
where the numerical values for the coefficients $f_{1,2,3}$ and $q_{1,2,3}$ are obtained by calibration with results from numerical simulations. 

Once the ringdown frequency and the quality factor are known, we can find the damping time of each QNM, using eq.\eref{eq:QtauQNM}.

\subsection{BoB versus the IRS}
\label{ssec:BoB-IRS}
After carefully building the stage by first grounding it at the LR and furnishing it with analytical expressions for the final spin, mass and QNM frequency, we are ready to introduce the two main candidates: the \emph{Backward-one-Body} (BoB) and the \emph{Implicitly-Rotating-Source} (IRS) models. 
Let's open the discussion between these two front-runners, aiming at determining which one will prevail, or if an agreement can be reached between them. 
We will restate the conceptual basis of those models, while referring the reader to the original papers for details on the mathematics.

Both approaches rely on NR simulations that solve numerically Einstein's equations for strong gravitational field, to model the highly nonlinear regime of the merger.
The computer simulations evolve the gravitational field as prescribed by the GR equations, by encoding the curvature of the geometry in the \emph{Riemann tensor} $R_{abcd}$, with indices $a,b,c,d=1...4$.  
This quantity is further separated into the \emph{Ricci tensor} $R_{ab}$ containing the curvature of the spacetime due to matter distribution, and the \emph{Weyl tensor} $C_{abcd}$ describing the curvature of the vacuum without matter and energy, but with non-zero gravitational field, when the spacetime is not flat.
Thus, the \emph{Weyl tensor} informs how the gravitational perturbation travels in empty space, and contains all the information needed to calculate the outgoing GW in the asymptotic limit.
To ensures that the extracted GW is gauge-free or independent of coordinates, the Weyl tensor is projected onto a vector basis called \emph{complex null tetrad} $(l,n, m,\bar{m})$ in the Newman-Penrose formalism \cite{newman}.
Here $(l,n)$ are real \emph{light-like} (null) inward and outward radial vectors, and $(m, \bar{m})$ is a  complex vector mapping a 2-sphere.
This projection yields the outgoing \emph{Weyl scalar} $\Psi_4 = C_{abcd} n^a \bar m ^b n^c \bar m^d$, a key term in the numerical calculation of the GW, which is related to the strain by the formula
 \begin{equation}
\label{eq:Psi4}
\Psi_4(t) = \frac{\partial^2 h_{NR}(t)}{\partial^2 t}.
\end{equation} 
The numerical relativistic strain $h_{NR}(t)$ is obtained by integrating eq.\eref{eq:Psi4} twice in time, assuming that the Weyl scalar is well described by a wave function of the form: 
$\Psi_4(t) = \mathcal{A}(t) e^{-i \phi_{GW}(t)}$, where $\mathcal{A}(t)=|\Psi_4(t)|$ and $\phi_{GW}(t)$ is the phase of the GW. 
This integration yields a proportionality relation of the form \cite{arXiv:1006.1632}:
\begin{equation}
\label{eq:strain}
h_{NR}(t) \propto -\frac{|\Psi_4(t)|}{\omega_{GW}(t)^2}e^{-i\phi_{GW}(t)} = -\frac{\Psi_4(t)}{\omega_{GW}(t)^2}.
\label{strainPsi4}
\end{equation}

\subsubsection{The IRS model}
\label{sssec:IRS}
Introduced in \cite{arXiv:0805.1428}, this model considers that the merger can be modeled as a single perturbed black hole that generates the GW \emph{implicitly} through its rotation, wherefore the name of \emph{Implicit Rotating Source}.
The GW thus emitted are approximated with damped sinusoids orbiting the newly formed black hole at the LR, with frequency and amplitude depending on the mass and the spin of this final black hole after it settles down. 

This model was generalized to include the ring-down phase at the end of the merger (gIRS) and was used in a series of papers to build complete GW waveforms for dynamical capture binaries \cite{arXiv:1212.0837}, neutron star-black hole mergers \cite{arXiv:1403.0561}, and eccentric binaries \cite{arXiv:1403.7754, arXiv:1609.05933, arXiv:1711.06276}.

In short, this model employs a hyperbolic tangent to analytically describe the evolution of the orbital mode frequency during the merger and ringdown, of the form:
\begin{equation}
\label{eq:omTIRS}
\Omega_{IRS}(t) = \Omega_{i} +(\Omega_{QLM} - \Omega_{i})\left ( \frac{1}{2} (1 + \tanh [\ln{\sqrt{k}} + {\frac{t-t_{0}}{\tau}}]) \right )^k, 
\end{equation}
In this formula $\Omega_i$ is the initial frequency, corresponding to the inspiral frequency in the transition region, beyond the innermost stable circular orbit (ISCO), located near the light ring for comparable mass binaries \cite{arXiv:0706.3732}.
Moreover, $\tau$ is the damping time of the QNM from eq.\eref{eq:QtauQNM}, $t_{0}$ is the time when the strain rate of the wave reaches its peak amplitude, and $k$ is a free parameter that controls the shape of this analytic expression, found through an heuristic fit to a large number of simulated GW waveforms \cite{arXiv:0805.1428}. 
The initial frequency $\Omega_{i}$ is further replaced with the formula \cite{arXiv:0805.1428}:
\begin{equation}
\label{eq:omIRS}
\Omega_i = \Omega_{QLM} -\frac{\tau}{2} \dot \Omega_0 \left (1 + \frac{1}{k} \right )^{1+k}, 
\end{equation}
such that the parameter $k$ accounts for the initial frequency in this model.
Here, $\dot \Omega_0$ is the value of the time derivative in frequency at $t_0$.
With this expression, eq.\eref{eq:omTIRS} is further rewritten as depending only on the frequency of the orbital QNM in  \cite{arXiv:1107.1181}:
\begin{equation}
\Omega_{IRS}(t)= \Omega_{QNM} (1 - f_{IRS}(t) )
\label{eq:omegaIRS}
\end{equation}
where the analytic function $f(t)$ takes the form:
\begin{equation}
f_{IRS}(t)= \frac{c}{2} \left (1+\frac{1}{k} \right )^{1+k}  \left [1 - \left (1 + \frac{1}{k}e^{-2(t-t_0)/\tau} \right )^{-k} \right ],
\label{eq:fIRS}
\end{equation}
with $c$ another free parameter, given by $c = \dot \Omega_{0} \tau/\Omega_{QNM}$ in \cite{arXiv:1107.1181} and tuned to NR data.

The frequency given by eq.\eref{eq:omegaIRS} is integrated to find the phase of the dominant GW mode during the merger:
\begin{equation}
\phi_{IRS}(t) = \int 2 \Omega_{IRS}(t) dt =  \int \omega_{IRS}(t) dt.
\label{eq:phiIRS}
\end{equation}
With the evolution of the frequency in hand, we turn now to the amplitude of the wave. 
Using the proportionality between the energy flux of the GW and its squared amplitude deduced in \cite{arXiv:0805.1428}, we can write for the strain-rate amplitude the expression: 
\begin{equation}
A_{\dot h}(t)  = \sqrt{16\pi \dot{E}(t) }. 
\label{eq:A2IRS}
\end{equation}
In the IRS model, the source generates GW by losing its rotational energy, which is proportional to the square of the orbital frequency, thus its time derivative will be:
\begin{equation}
\dot{E}(t)  = \xi \Omega(t)  \dot{\Omega}(t),
\label{eq:EdotIRS}
\end{equation}
where the parameter $\xi = \partial J(t) / \partial \Omega(t) $ measures the change in the angular momentum with the orbital frequency. 
Assuming conservation of the angular momentum, this parameter becomes constant and can be determined by imposing its continuity close to the LR. 
It is shown in \cite{arXiv:1107.1181, arXiv:1403.7754} that a good approximation for the amplitude of the dominant mode of the strain is given by:
\begin{equation}
A_{IRS}(t)=\frac{A_{0,IRS}}{ \omega_{IRS}(t)} \left (\frac{|\dot f_{IRS}(t)|}{1 + \alpha (f_{IRS}(t)^2-f_{IRS}(t)^4)}\right )^{1/2},
\label{eq:ampIRS}
\end{equation}
where $A_{0,IRS}$ is an amplitude scale factor and  $\alpha$ is a third free parameter chosen again by tuning the model to numerical simulations \cite{arXiv:1212.0837, arXiv:1609.05933}.

Once the phase eq.\eref{eq:phiIRS} and the amplitude eq.\eref{eq:ampIRS} of the GW are determined, the strain is easily formed with the analytical expression:
\begin{equation}
h_{IRS}(t)  = A_{IRS}(t) e^{-i\phi_{IRS}(t) }.
\label{eq:strainIRS}
\end{equation}
This model depends on the following three free numerical parameters required to construct the strain of the GW: $\alpha, k$ and $c$, determined from fits with NR. 
Numerical values for those parameters are provided in \cite{arXiv:1107.1181, arXiv:1212.0837, arXiv:1609.05933}. 
As initial data we must know only the QNM frequency $\Omega_{QLM}$. 
This model does not attempt to give any information on the initial time $t_i$, which is a free fitting parameter as well, picked around the LR.

\subsubsection{The BoB Model}
\label{sssec:BoB}
Introduced in \cite{arXiv:1810.00040}, this model approaches the merger dynamics in a backward fashion, by starting from the perturbed final black hole  and building the GW back in time, to arrive at the LR, wherefore the name of \emph{Backward-one-Body}. 
In this picture, the GW are emitted by the fictitious point-mass $\mu$, called \emph{perturber}, who loses orbital energy as its orbit shrinks, until it falls onto the stationary massive body of mass $M=m_1+m_2$. 
As the perturber inspirals emitting GW, this radiation reflects off the LR, providing thus the majority of signal.
The ringdown starts once the perturber passes through the LR, and from that point on, most of its radiated GW falls back onto the massive black hole and thus can no longer be observed. 
The GW perturbations emitted very close to the LR, just before its crossing, have higher frequencies and are trapped there longer in closed \emph{null orbits}, as it was the case with the photons. 
This high-frequency radiation, reflected and trapped at the LR and then escaping at later time, is what we observe during and even after the merger, as the QNM of the black hole ringdown.
This GW signal can be modeled analytically as exponentially decaying sinusoids that break free from the LR on null geodesics \cite{mashhoon}. 
By tracing those waves back in time to the point where the null geodesics meet again, we can deduce the behavior of the signal at earlier times.

We sketch below a summary of the BoB ansatz following  \cite{arXiv:1810.00040}, which starts with modeling the trajectory of the perturber around the LR with the equation:
\begin{equation}
r(t) = r_{LR}[1+\epsilon f(t-t_0)],
\label{eq:rpert}
\end{equation}
where $r_{LR}$ marks the position of the LR, $\epsilon$ is a small scaling factor and $t_0$ is again the time when the gravitational waves reach their peak amplitude. 
The shape of the perturbation function is chosen to be: 
\begin{equation}
f(t) = \sinh\left(\frac{t}{\tau}\right).
\label{eq:fpert}
\end{equation}

We know from geometric optics that in the eikonal approximation the amplitude of the energy carried by the waves along the null rays satisfies the transport equation \cite{book:optics}:
\begin{equation}
\frac{d}{dt}(dr A(t) ) = 0 
 \label{eq:transport}
\end{equation}
By taking the time derivative of eq.\eref{eq:rpert} we obtain $dr \propto \cosh\left(\frac{t-t_p}{\tau}\right)$, which correctly describes the radial spreading of the gravitational perturbation. 
If we replace $dr$ with its $\cosh$ dependency in eq.\eref{eq:transport} and integrate it, we obtain the amplitude as:
\begin{equation}
A (t)= A_{0,BoB} \mbox{sech} \left(\frac{t-t_0}{\tau}\right),
\label{eq:ampBoB}
\end{equation}
where $A_{0,BoB}$ is again an integration constant that enters as a scaling factor. 

In the BoB model, the amplitude of the perturbations is taken to correspond to the Weyl scalar $A(t)  = |\Psi_4(t) |$.
According to eq.\eref{eq:strain}, and assuming that the proportionality coefficients remain approximately constant or change much slower than the phase during the merger, the amplitude of the GW strain becomes:
\begin{equation} 
|h_{BoB}(t) | =  \frac{|\Psi_4(t)|}{ \omega_{GW}(t)^2}  = \frac{A(t)}{m^2 \Omega_{BoB}^2(t)^2} 
 \label{eq:ABoB}
\end{equation}
where, as before, $\Omega_{BoB}(t)$ denotes the frequency of the orbital mode, and $m=2$ for the dominant mode. 
We proceed next with calculating of frequency from this amplitude. 
Using the insights given by eqs. \eref{eq:A2IRS} and \eref{eq:EdotIRS} from the IRS model, we find a relationship between the amplitude and frequency of the orbital GW mode:
\begin{equation}
\mbox{sech}^2\left(\frac{t-t_0}{\tau}\right) = 16\pi \xi \Omega(t)^3 \dot \Omega(t).
\label{eq:AOmg3}
\end{equation}
Noting that $\xi$ stays constant during the merger timescale given by the damping or \emph{e-folding} time $\tau$,  a constant $\kappa$ is introduced to encapsulate this proportionality: $  \kappa = \tau/(4\pi \xi)$ \cite{arXiv:1810.00040}. 

Subsequently, eq.\eref{eq:AOmg3} is integrated to obtain the frequency as: 
\begin{equation}
\Omega_{BoB}(t) = \left( \Omega_i^4+\kappa \left[ \tanh \left(\frac{t-t_0}{\tau}\right) - \tanh \left(\frac{t_i-t_0}{\tau}\right) \right]\right)^{1/4}.
\label{eq:omgBoB}
\end{equation}
Now, the BoB model detaches from the IRS model, because here the constant $\kappa$ is no longer a free parameter, but is constrained by integrating eq.\eref{eq:AOmg3} written as:
\begin{equation}
\frac{\kappa}{4 \tau}\mbox{sech}^2\left(\frac{t-t_0}{\tau}\right) dt = \Omega(t)^3 d \Omega(t),
\label{eq:kAOmg3}
\end{equation}
between $\Omega(t \rightarrow t_i) = \Omega_{i}$ and $\Omega(t \rightarrow \infty) = \Omega_{QNM}$ instead of fitting it to the numerical data, as in the case of the IRS model.
Thus, the constant $\kappa$ becomes determined by:
\begin{equation}
\label{eq:constk}
\kappa = \left[ \frac{\Omega_{QNM}^4-\Omega_i^4}{1 -\tanh\left(\frac{(t_i-t_0)}{\tau} \right)}\right]
\end{equation}
Once the analytical form of this $\kappa$ constant is known, we substitute it back into eq.\eref{eq:omgBoB} and calculate analytically the initial time $t_i$ corresponding to the matching frequency $\Omega_i$ between the end of the inspiral and the beginning of the merger:
\begin{equation}
t_i = t_0 - \frac{\tau}{2}\ln \left( \frac{ \Omega_{QNM}^4 - \Omega_i^4}{2\tau\Omega_i^3\dot{\Omega}_i} - 1 \right). 
\label{eq:tisimp}
\end{equation}

With the expressions for $\kappa$ and $t_i$ calculated, we return now to the key equations \eref{eq:ampBoB} and \eref{eq:omgBoB} to build the strain. 
We obtain first the phase by integrating eq.\eref{eq:omgBoB}. 
\begin{equation}
\phi_{BoB}(t) = \int_{t_i}^{t_f} 2\Omega_{BoB}(t) dt = \int_{t_i}^{t_f} \omega_{BoB}(t) dt.
\label{eq:phiBoB} 
\end{equation}
Here, $t_f$ is left as a free choice, but $t_i$ is fixed by eq.\eref{eq:tisimp}. 
To calculate the amplitude of the strain, we will need to use eq.\eref{strainPsi4}, because eq.\eref{eq:ampBoB} refers to the amplitude of the Weyl scalar $\Psi_4$. 
The expression for the BoB strain will read:
\begin{equation}
h_{BoB}(t) = -\frac{A(t)}{\omega_{BoB}(t)^2}e^{-i\phi_{BoB}(t)}.
\label{eq:strainBoB}
\end{equation}

The obvious advantage of the BoB model is that it does not depend on any free parameters obtained from direct fit with numerical simulations data.
This model still depends indirectly on NR results, through the formulas used to determine the mass, spin and QNM frequency of the final black hole.   

\section{Results}
\label{sec:results}

\subsection{Before the Beginning}
\label{ssec:before}

Before we go on with the implementation of the merger models, we begin with the end in mind, because we must first calculate the final spin, mass and quality factor of the remnant black hole. 
For the spin we use first eq.\eref{eq:chifin} with parameters for $s_{ij}$ from \cite{arXiv:0710.3345, arXiv:0712.3541}, as detailed in \eref{eq:chifin1}. 
We implement as well alternate forms of the final spin as given in \cite{arXiv:1406.7295, arXiv:1610.09713, arXiv:1611.00332} and presented them in \ref{appendixA}.

We implicitly assume only the components of the individual spins along the angular momentum, and find that increasing the mass-ratio while keeping the spins equal will lower the final spin, while increasing the spin of the heavier black hole will increase the final spin as well. 

With the final spin known, we proceed next to calculate the mass loss to GW energy, which subtracted from the total mass gives the mass of the final black hole, an essential quantity for the merger models, as explained in Sub\sref{sssec:mass}.
Eq.\eref{eq:MChris}, although exact and accurate, does require the knowledge of the apparent horizon, which is used to mark the boundary of the distorted black hole formed by collision, and it can only be determined numerically \cite{arXiv:1703.00118}, although recently it was evaluated using the available GW events \cite{arXiv:2202.08848}.
To circumvent this problem, we implement five analytical formulas for estimating $M_f$, based on polynomial fits.
The first three expressions are for equal-mass, spin-aligned black holes, and we transcribe them in \ref{appendixB}, eq.\eref{eq:MfAEI}, eq.\eref{eq:MfRIT} and  eq.\eref{eq:MfRIT1}. 
The next two formulas are extended to include unequal masses, as specified in eq.\eref{eq:MfRIT2} and \eref{eq:MfH}.  
We take the average of the final mass estimated by all those analytical approximations and use it as initial data. 
For unequal masses we consider only eqs.\eref{eq:MfRIT2} and \eref{eq:MfH}. 

The analytical dependence of the final mass on the spins and mass ratios shows that the final mass increases with mass ratio and decreases with the increase of the spins. 
Thus, highly spinning, equal mass binaries will release more energy in GW, making those events easier to be detected than unequal-mass, slowly spinning systems.

Two other key elements that we must add to the initial data of the merger models are the knowledge of the ringdown frequency $\Omega_{QNM}$ and of the quality factor $Q_{QNM}$, which is necessary to calculate the damping or \emph{e-folding} time describing the decay in the amplitude of the GW as the remnant black hole settles down.
We calculate the frequency of the dominant $(l = m = 2)$ QNM mode and the quality factor using eqs.\eref{eq:MfOQNM} and \eref{eq:QQNM}, with the coefficients $(f_i, q_i)$ from \cite{echeverria, arXiv:0905.2975} and verify these results against the values given in Table VIII of \cite{arXiv:gr-qc/0512160}.
We obtain the orbital frequency as $\Omega_{QNM} = \omega_{22}/2$, and find its damping time from eq.\eref{eq:QtauQNM}.

After implementing those formulas, we calculate the mean values for the final mass $M_f$, spin $\chi_f$, dominant mode of the resonant frequency $\omega_{QNM}$, and corresponding quality factor $Q_{QNM}$ for three non-spinning binary systems with mass ratios $q =1, 2, 4$.

The last essential ingredient is the orbital angular frequency of the binary around ISCO, where the transition from the inspiral to merger is located.
This frequency is required by the BoB model, but is implicitly given in the IRS model by the constant $k$.
We use for this frequency the expression given in \cite{bardeen, arXiv:gr-qc/0001013}:
\begin{equation}
\omega_{i} = \frac{1}{{\tilde r_{i}^{3/2}} + \chi_{f}},
\label{eq:omISCO}
\end{equation}
where 
\begin{equation}
{\tilde r_{i}} = 3 + Z_2 - \sqrt{(3 - Z_1)(3 + Z_1 + 2 Z_2)} ,
\label{eq:rISCO}
\end{equation}
with $Z_1= 1 + (1 - \chi_f^2)^{1/3}[(1 + \chi_f)^{1/3}+ (1 - \chi_f)^{1/3}]$ and $Z_2 = \sqrt{3 \chi_f^2 + Z_1^2}$.
We also need the slope of this initial frequency $\dot \Omega_{i}$, for which we do not have an analytical expression. 
Instead, we make an informed choice for this quantity, corresponding to the value for $\Omega_{i}$ from our previous post-Newtonian implementation (see \cite{arXiv:2203.08998}).
Table \ref{tab:mergeID} contains the calculated mean values for the final spin and mass, the ringdown frequency, quality factor and damping time, together with the initial frequency and its derivative, used as initial data necessary to implement the BoB and IRS models.
\small
\begin{table}[!htbp]
\caption{\label{tab:mergeID}{Initial data required for comparing the merger models BoB and IRS.}}
\begin{indented}
\item[]\begin{tabular}{@{}llllllll}
\br
$q$ & $\chi_f$   & $M_f$      & $\omega_{QNM}$ & $Q_{QNM}$ & $\tau$          & $\Omega_i$ & $\dot \Omega_i$\\
\mr
$1$ & $0.6865$ & $0.9516$ & $0.5698$               & $3.301$      & $11.586$   &$7.529\times10^{-2}$ & $8.577\times10^{-4}$\\
\mr
$2$ & $0.6231$  & $0.9612$ & $0.5385$              & $3.056$      & $11.349 $   &$7.414\times10^{-2}$ & $7.399\times10^{-4}$  \\
\mr
$4$ & $0.4637$  & $0.9779$  & $0.4782$             & $2.642$      & $11.049$   &$7.169\times10^{-2}$ & $5.046\times10^{-4}$\\
\br
\end{tabular}
\end{indented}
\label{tbl:IDBoBvalues}
\end{table}
\normalsize

\subsection{Playing in Tune}
\label{ssec:BoBcomp}
With the initial data in hand, we are ready to compare against each other the two analytical approximations for calculating the GW emitted during the merger: BoB and IRS, and to gauge their performance against the numerical data. 
From the large database of numerical waveforms provided by the SXS catalog for equal-mass non-spinning BBH, we choose SXS:BBH:0180 for our comparison.

As emphasized above, the implementation of the IRS model does not require the knowledge of the frequency at ISCO, relying instead on direct fits to NR simulations to estimate the necessary constants. 
We take those parameters from \cite{arXiv:1609.05933} that gives only the fits for non-spinning binaries, which limits our comparison to those mergers. 

\begin{figure}[!ht]
\centering
\includegraphics[scale=0.75]{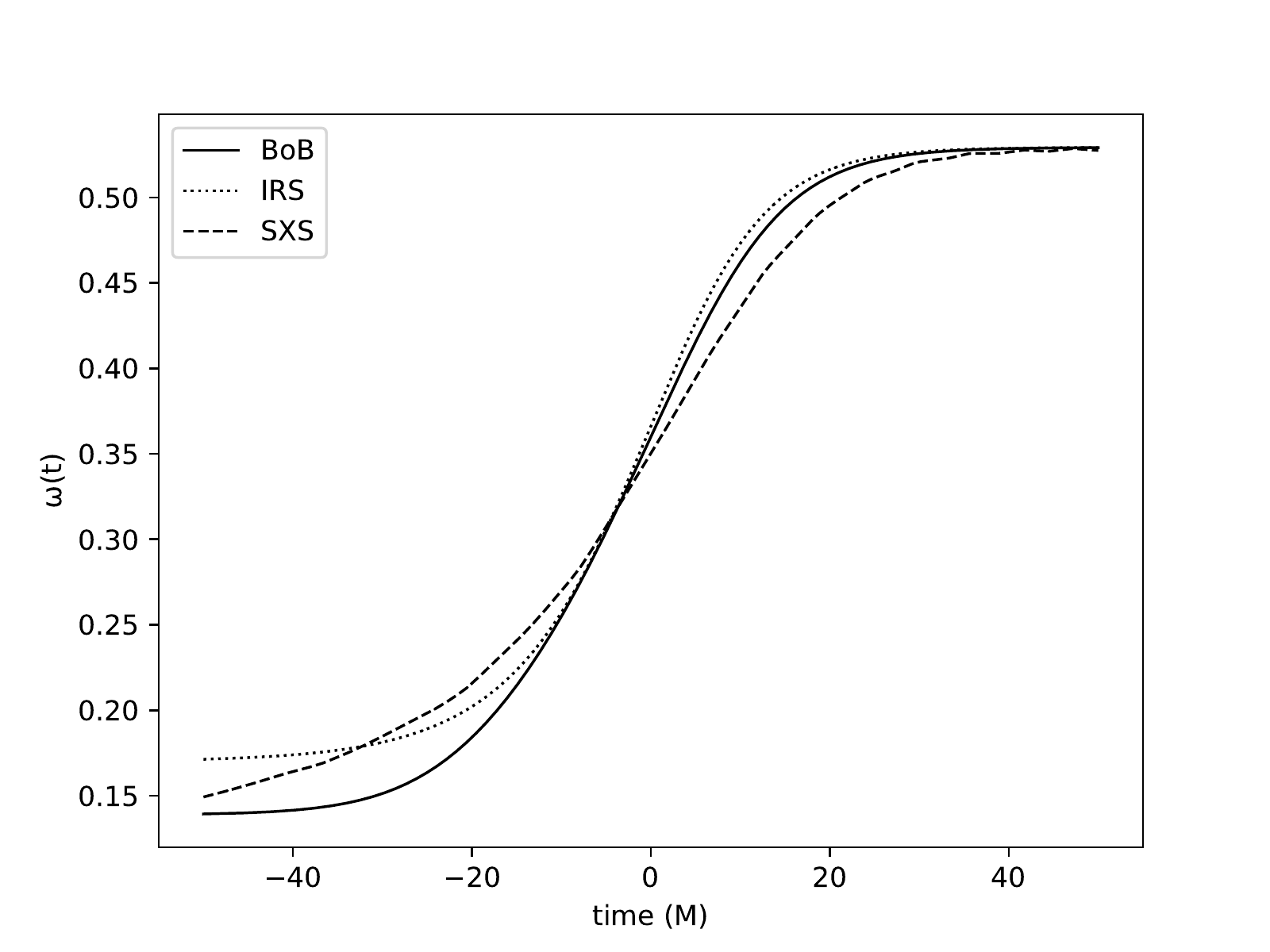}
\caption{Comparison between the analytical and numerical frequencies, aligned at the peak of increase in frequency, for an equal-mass non-spinning BBH merger. The BoB frequency was translated with $t_f$.} 
\label{fig:wBoBIRSComp}
\end{figure}

Let's proceed by calculating first the angular frequency of the GW.
For  the IRS model this is given by eq.\eref{eq:omegaIRS} with the function $f(t)$ from eq.\eref{eq:fIRS} and the numerical values for the pair of constants $(c,k)$ as detailed in \cite{arXiv:1609.05933}. 
For the BoB model we start by calculating the angular frequency of the orbital mode using eq.\eref{eq:omgBoB}, with $\kappa$ given by eq.\eref{eq:constk} and $t_i$ from eq.\eref{eq:tisimp}, and obtain the dominant mode by multiplying with $m=2$. 
We compare those frequencies by translating them to overlap at the peak of the increase in their rate.
This time coincides with the peak in amplitude for IRS, while for BoB it occurs earlier, which is the expected behavior \cite{arXiv:0805.1428}.
We denote this time by $t_f$.
This results into a remarkably good overlap in frequency for both models near, during and after the merger time, as seen in Fig.\ref{fig:wBoBIRSComp},
with an expected discrepancy around the initial ISCO frequency. 
This is caused by the fact that the IRS model hard-codes the initial frequency from fits with numerical data and seems to overestimate its value. 	
We also implement eqs.\eref{eq:omTIRS} for comparison, but obtain a less accurate fit with the angular frequency of the BoB model.
The comparison with the numerical data shows that the analytical models predict a slightly steeper increase in frequency around the merger.

\begin{figure}[!ht]
\centering
\includegraphics[scale=0.75]{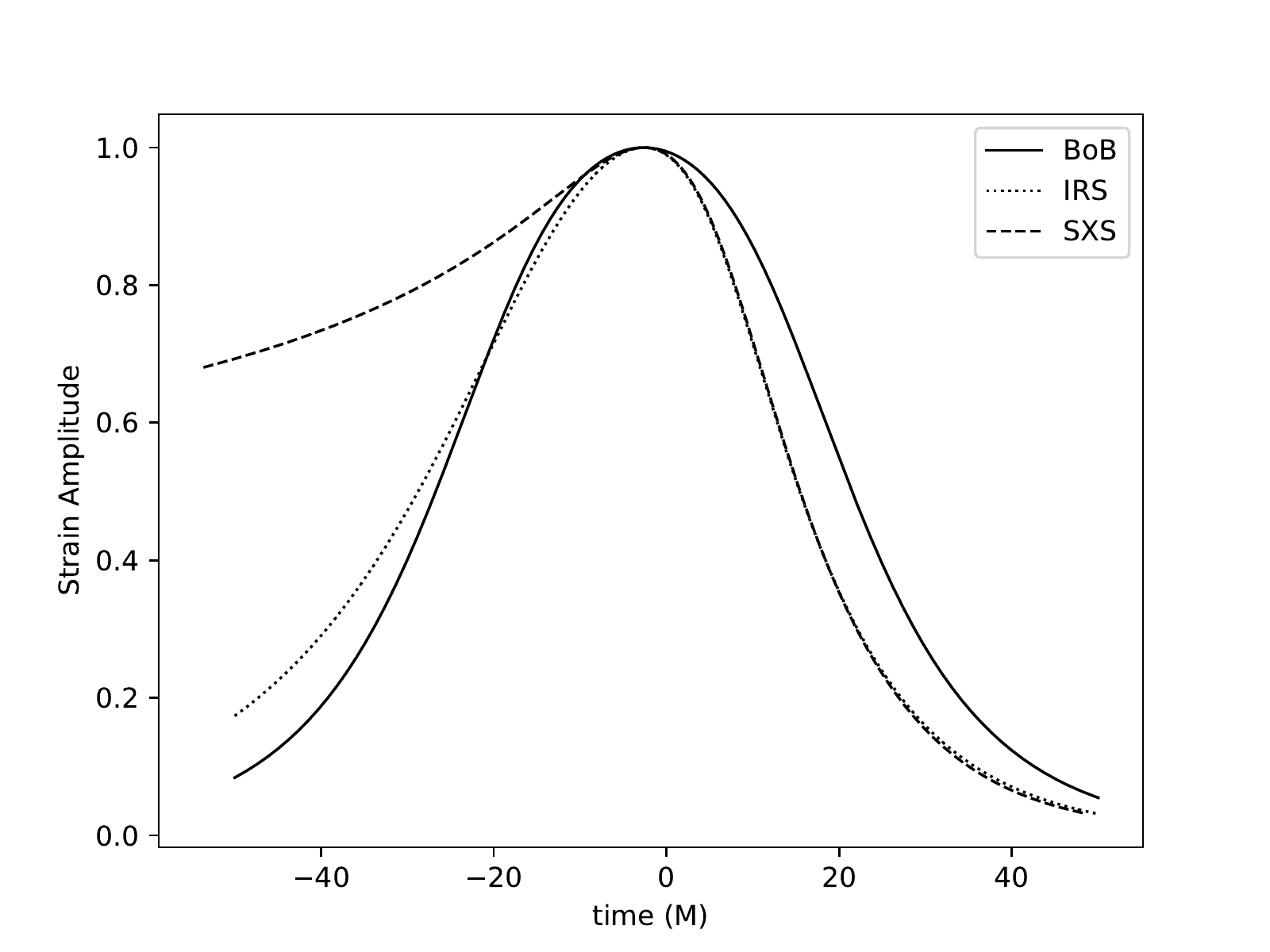}
\caption{Comparison between the normalized analytical and numerical amplitudes aligned at the peak, for an equal-mass non-spinning BBH. The BoB amplitude was translated with $t_p$.} 
\label{fig:hAmpBoBIRSComp}
\end{figure}

Once the frequency is known, we calculate the amplitude, given for the IRS model by eq.\eref{eq:ampIRS} with the numerical constant $\alpha$ as in \cite{arXiv:1609.05933} and for the BoB model by eq.\eref{eq:ampBoB}. 
For comparison, we fit the amplitudes at their peak, which for IRS coincides with $t_0 = 0$, while for the BoB model it is again displaced slightly, and we denote this time by $t_p$.
The amplitude of the Weyl scalar peaks indeed at $t_0 = 0$ in this model, while the amplitude of the strain, being scaled with the square in the frequency, reaches the peak earlier.  
The plot shown in Fig.\ref{fig:hAmpBoBIRSComp} reveals a noticeable discrepancy between the amplitudes given by the two analytical models, which is expected, considering the different formulas used for the evaluation of those amplitudes. 
The comparison with the numerical data favors the IRS amplitude, because IRS is a heuristic fit to numerical data. 
Both analytically evaluated amplitudes lose accuracy and deviate strongly from the NR-predicted value up to the LR, located around $t_{LR} \approx 6M$ \cite{bardeen}.

\begin{figure}[!ht]
\centering
\includegraphics[scale=0.75]{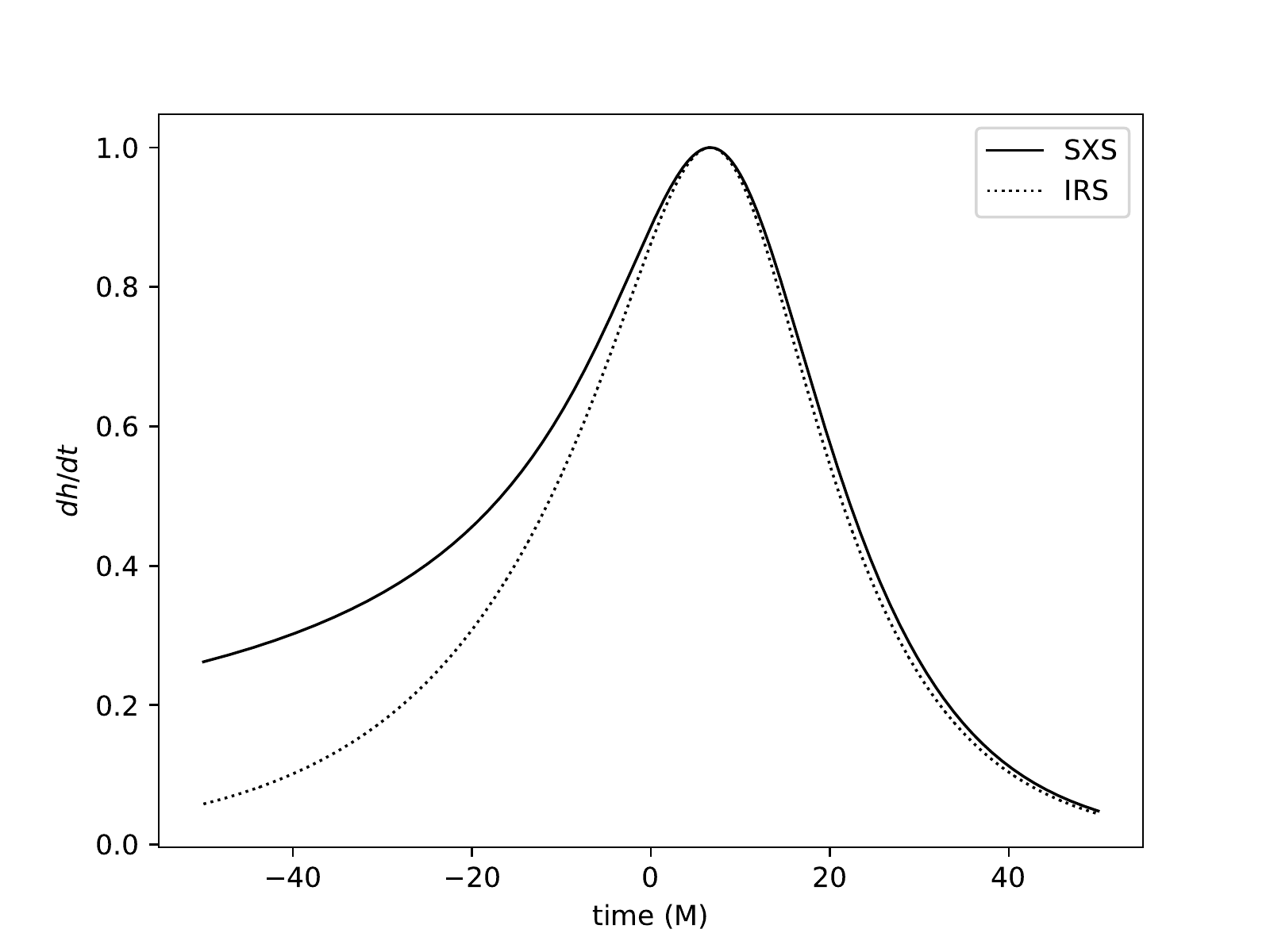}
\caption{Comparison between the normalized IRS analytical amplitude of the strain-rate and the numerical $|\dot h(t)|$ aligned at the peak, for an equal-mass non-spinning BBH.} 
\label{fig:dhAmpSXSIRSComp}
\end{figure}

\begin{figure}[!ht]
\centering
\includegraphics[scale=0.75]{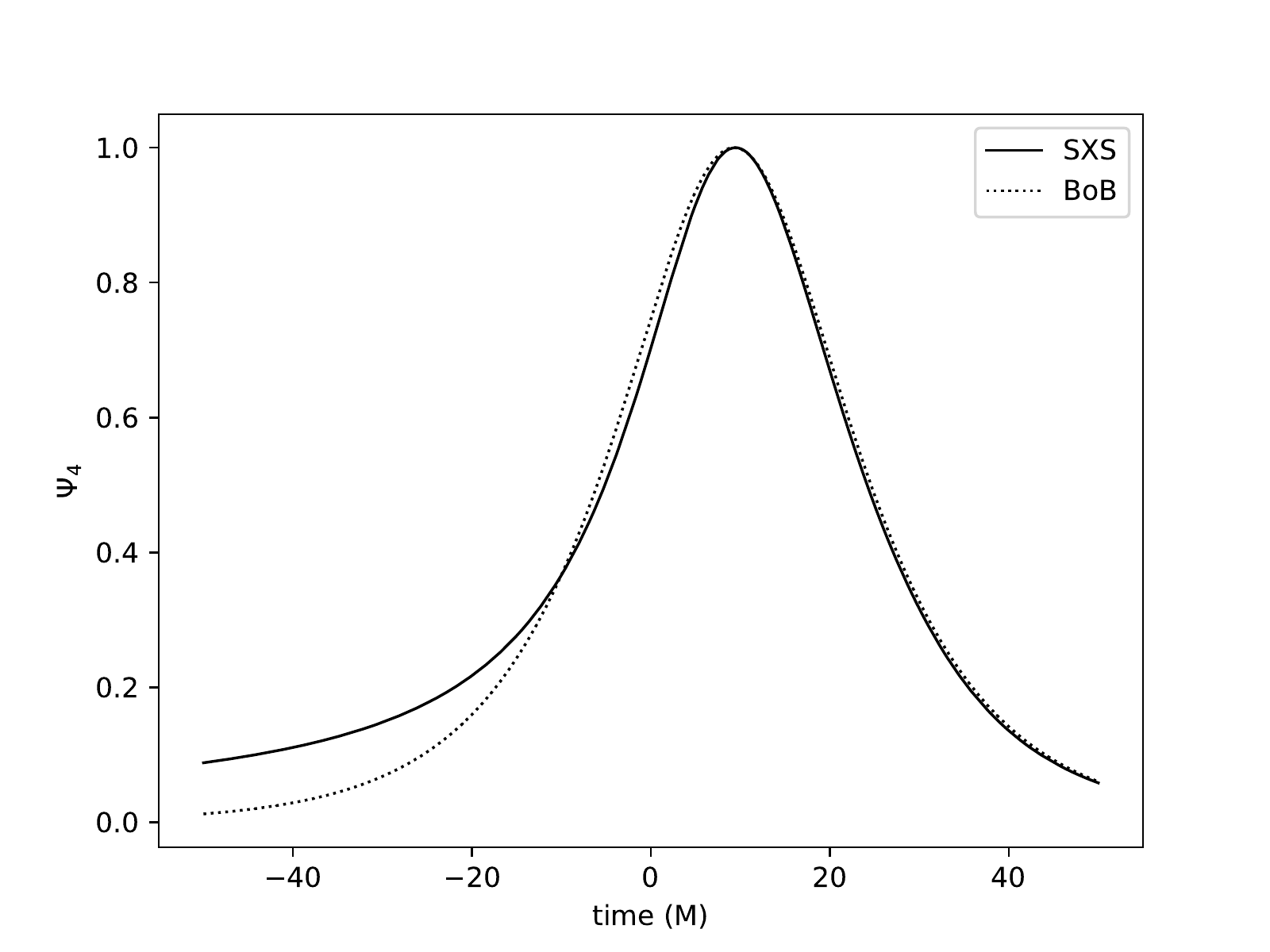}
\caption{Comparison between the normalized BoB analytical amplitude of $|\Psi_4(t)|$ and the numerical $|\Psi_4(t)|$ aligned at the peak, for an equal-mass non-spinning BBH.} 
\label{fig:ddhAmpSXSBoBComp}
\end{figure}

We note that we expect these discrepancies in the amplitudes of analytical models with respect to the numerical amplitude, due to the inherent assumptions built in the IRS and BoB models, which we test here.  
The amplitude used in the IRS model is built using the strain-rate amplitude $|\dot h(t)|$ from eq.\eref{eq:A2IRS}, and the amplitude of the BoB model from $|\Psi_4(t)|$ given by eq.\eref{eq:ampBoB}, by simply dividing by the waveform frequency, thus neglecting the changes in amplitude and frequency, and assuming that they change much slower than the phase during the merger and ringdown. 
We do mention that these extra terms that are not included in the analytical models will leave an imprint in the shape of the amplitude and will cause a time shift. 
To better asses the validity of those assumptions, we compare the amplitudes again as follows: in Fig.\ref{fig:dhAmpSXSIRSComp} we show the analytical amplitude of the strain-rate $|\dot h(t)|$ employed by the IRS model superimposed with the numerical $|\dot h(t)|$, and in Fig.\ref{fig:ddhAmpSXSBoBComp} we juxtapose the analytical $|\Psi_4(t)|$ proposed by BoB with the amplitude of the numerically calculated $|\Psi_4(t)|$.
Indeed, we see that the BoB analytical amplitude $|\Psi_4(t)|$ reproduces with better fidelity the numerical results than the $|\dot h(t)|$ amplitude of the IRS model.

\begin{figure}[!ht]
\centering
\includegraphics[scale=0.75]{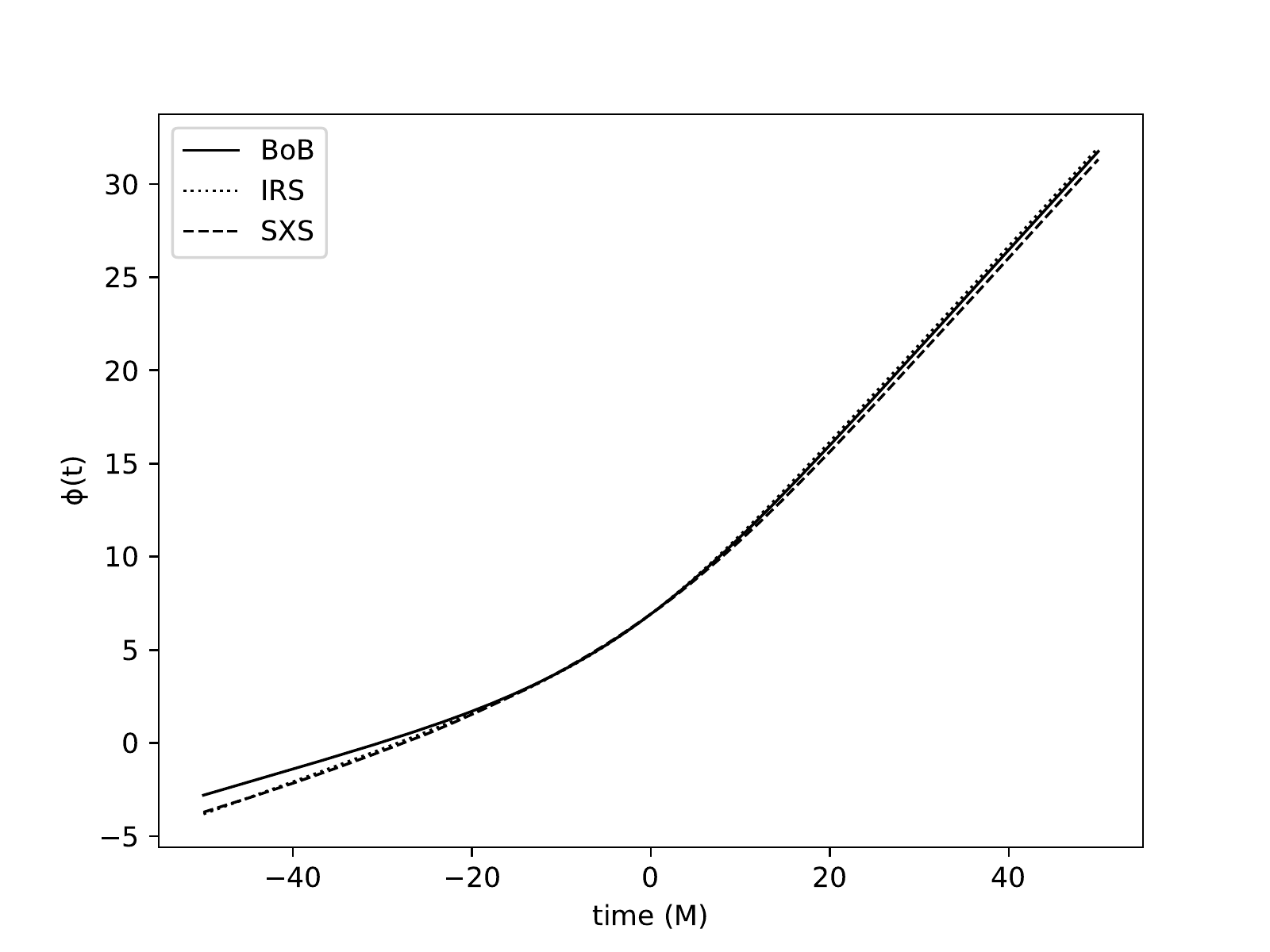}
\caption{Phase comparison, with $\phi_{IRS}$ corrected with the phase difference $\Delta t$ and $\phi_{BoB}$ translated with $t_p$ for an equal-mass non-spinning BBH.}
\label{fig:phiBoBIRSComp}
\end{figure}

We proceed next to calculate the phase, given for BoB by eq.\eref{eq:phiBoB} and for IRS by eq.\eref{eq:phiIRS}. 
We identify a phase difference between the two models at the peak in their frequency variation $\dot \Omega(t)$, and after correcting the IRS model for it, we translate $\phi_{BoB}$ with this time, which we denote it  $\Delta t$, obtaining an excellent fit, shown in Fig.\ref{fig:phiBoBIRSComp}.

Lastly we calculate the strain, given for the IRS model by eq.\eref{eq:strainIRS} and by eq.\eref{eq:strainBoB} for the BoB model. 
Due to the phase difference we uncovered, simply matching the two strains at their peak does not give the best overlap, even if we account for the phase difference when computing the strain of the IRS model with eq.\eref{eq:strainIRS}. 
Instead, our recourse is to employ the excellent match in frequency we found around the merger, and by translating the BoB model with the time of the peak in the frequency rate $t_f$, after we correct the IRS model with the phase difference, we obtain a very good fit, both in period and amplitude, of the GW strain predicted by the two models, and with the NR strain for the chosen data.

\begin{figure}[!ht]
\centering
\includegraphics[scale=0.75]{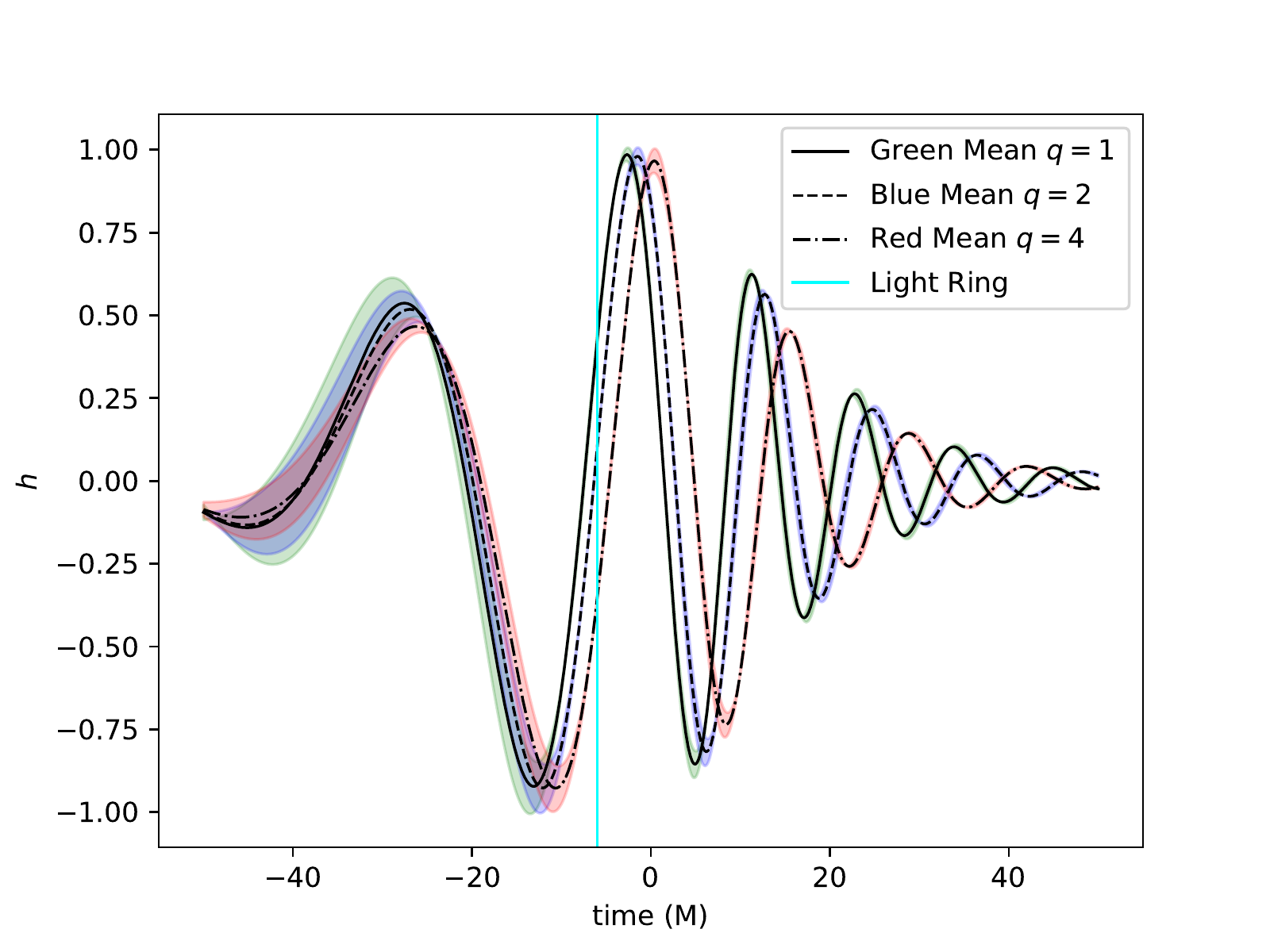}
\caption{Comparison between the BoB and phase corrected IRS for three mass ratios. The BoB strain was translated with $t_f$.
The vertical line marks the light ring.}
\label{fig:stdvmeanhplot}
\end{figure}

Let's examine now in more depth the consonance between the two models by comparing their strains, as described above, for three mass ratios $q= 1,2, 4$.
We plot in Fig.\ref{fig:stdvmeanhplot} the average between the two models and the standard deviation or the confidence interval for three mass rations $q=1,2,4$. 
We see that although the two models start at ISCO slightly out of sync due to the difference in their initial frequencies, they quickly get in accord with each other around the LR and continue in harmonious agreement around the peak and through the ringdown, for all three mass ratios.

\subsection{Spinning it Up}
\label{ssec:BoBspin}

We remind the reader that up to this point we have assumed that the black holes entering the merger do not spin, or their effective inspiral spin is zero, and indeed this seems to be the case for more than half the binary population detected in the LIGO and Virgo O1–O3 observing runs \cite{arXiv:2010.14533, arXiv:1912.11716}.  

In order to take into consideration individual spins when testing BoB, we will compare its predictions against a numerical GW template with spin, because it includes all the nonlinear physics, and is guaranteed to correctly model the merger. 
From the SXS catalog we choose the SXS:BBH:0746 simulation of an equal mass binary with large individual spins, for a direct comparison with the BoB model, both to test our code, and to confirm the accuracy of the BoB model at high spins. 

We collect the dimensionless spins of each BH from the SXS simulation, namely $\chi_1 = 0.5779$ and $\chi_2 = 0.7978$, then use our implementations for the final spin and mass as explained in \ref{ssec:before}. 
Our calculation of the final spin gives $\chi_{f} = 0.8806\pm0.0008$, which is very close to the value given by the SXS simulation $\chi_{f, NR}=0.8807$.
Our result for the final mass is slightly biased towards a larger mass remnant: $M_f = 0.9223\pm 0.0046$, comparing to the final mass from the SXS simulation $M_{f,SXS} = 0.9187$.
Next we generate the QNM frequency and quality factor analytically, as explained in Sub\sref{ssec:before}. 
The only two quantities left to determine are the initial  $\Omega_i$ and its slope $\dot \Omega_i$.
We calculate the initial frequency using eq.\eref{eq:omISCO} and denote it by $\omega_{i}$, with $\Omega_{i} = \omega_{i}/2$.
We also implement the fit for the initial frequency used in \cite{arXiv:1810.00040}, which we give in eq.\eref{eq:OmBoBisco}, denote it by $\Omega_{ISCO}$, and use it as it is. 
\begin{equation}
\label{eq:OmBoBisco} 
\Omega_{ISCO} = \frac{-0.091933 \chi_f + 0.097593}{\chi_f^2 - 2.4228 \chi_f + 1.4366}.
\end{equation}
Again, we do not calculate $\dot \Omega_i$, but use the same value provided in Table \ref{tab:mergeID} for the equal-mass binary, choice informed by our post-Newtonian evolution \cite{arXiv:2203.08998}. 
We recognize this ambiguity introduced by the choice in the time derivative of the  initial frequency as the highest source of error in our implementation. 
We will show below the effect of the choice in the initial frequency on the shape of the signal close to and after the merger, which tests the dependence of the BoB model on the initial data.

\begin{figure}[!ht]
\centering
\includegraphics[scale=0.75]{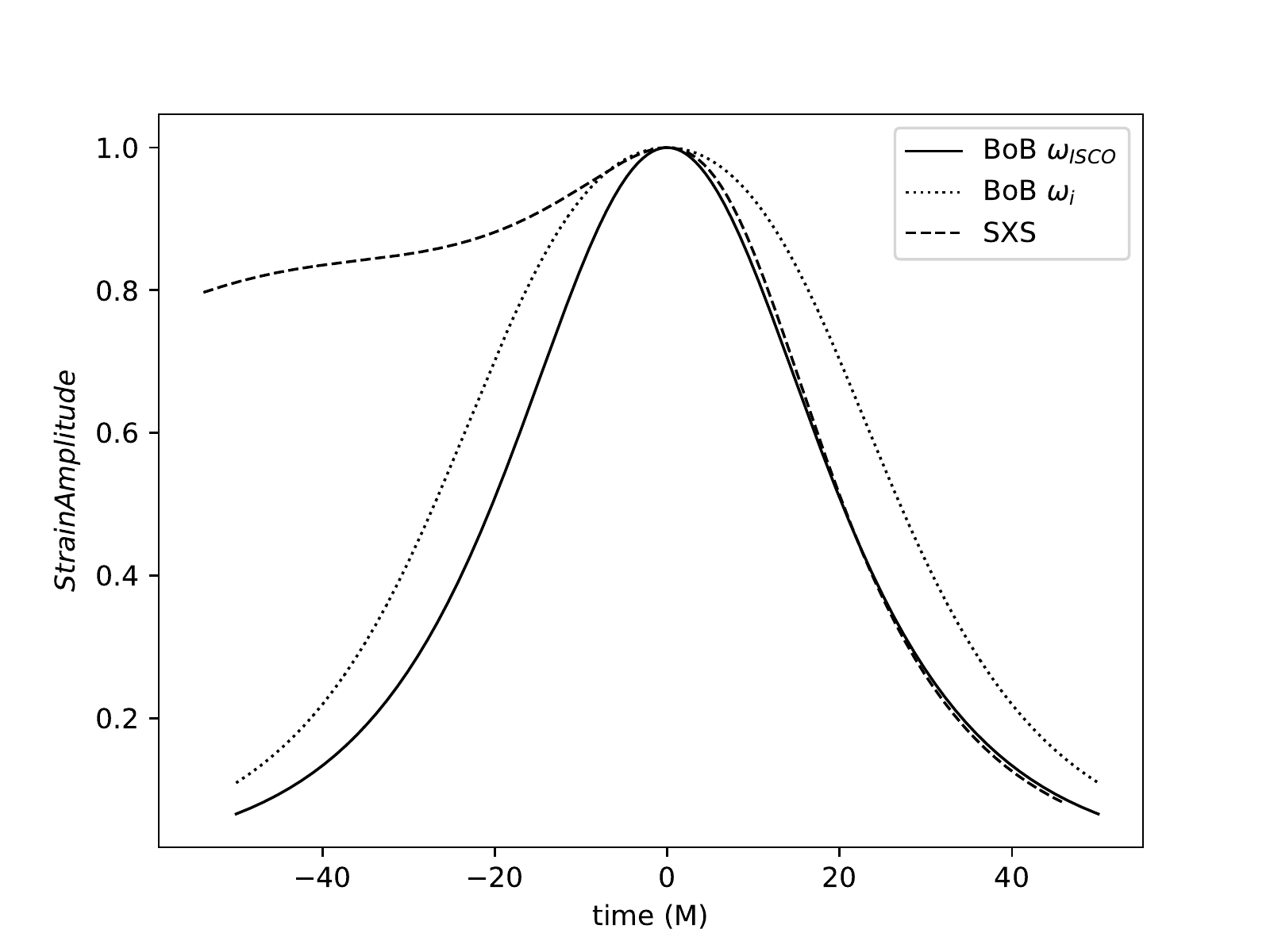}
\caption{The BoB amplitude for two initial frequencies, compared with the SXS amplitude, aligned at the peak, for an equal-mass BBH with $\chi_f = 0.88$.}
\label{fig:AmpBoBSXScomp}
\end{figure}

We start our comparison with the amplitudes, by first normalizing and aligning them at the peak, as is shown in Fig.\ref{fig:AmpBoBSXScomp}. 
The BoB amplitude calculated using the ISCO frequency choice has a smaller spread and exhibits a remarkable overlap with the SXS amplitude from the peak onwards, while the BoB amplitude using $\Omega_i$ as initial frequency has an overall larger spread and a weaker concordance with the shape of the SXS amplitude, even after the merger, similar with what we found in the non-spinning case, when we took into account IRS as well.
In both cases, the amplitude of the BoB model reveals a large discrepancy with the SXS predicted amplitude before the LR. 

\begin{figure}[!ht]
\centering
\includegraphics[scale=0.75]{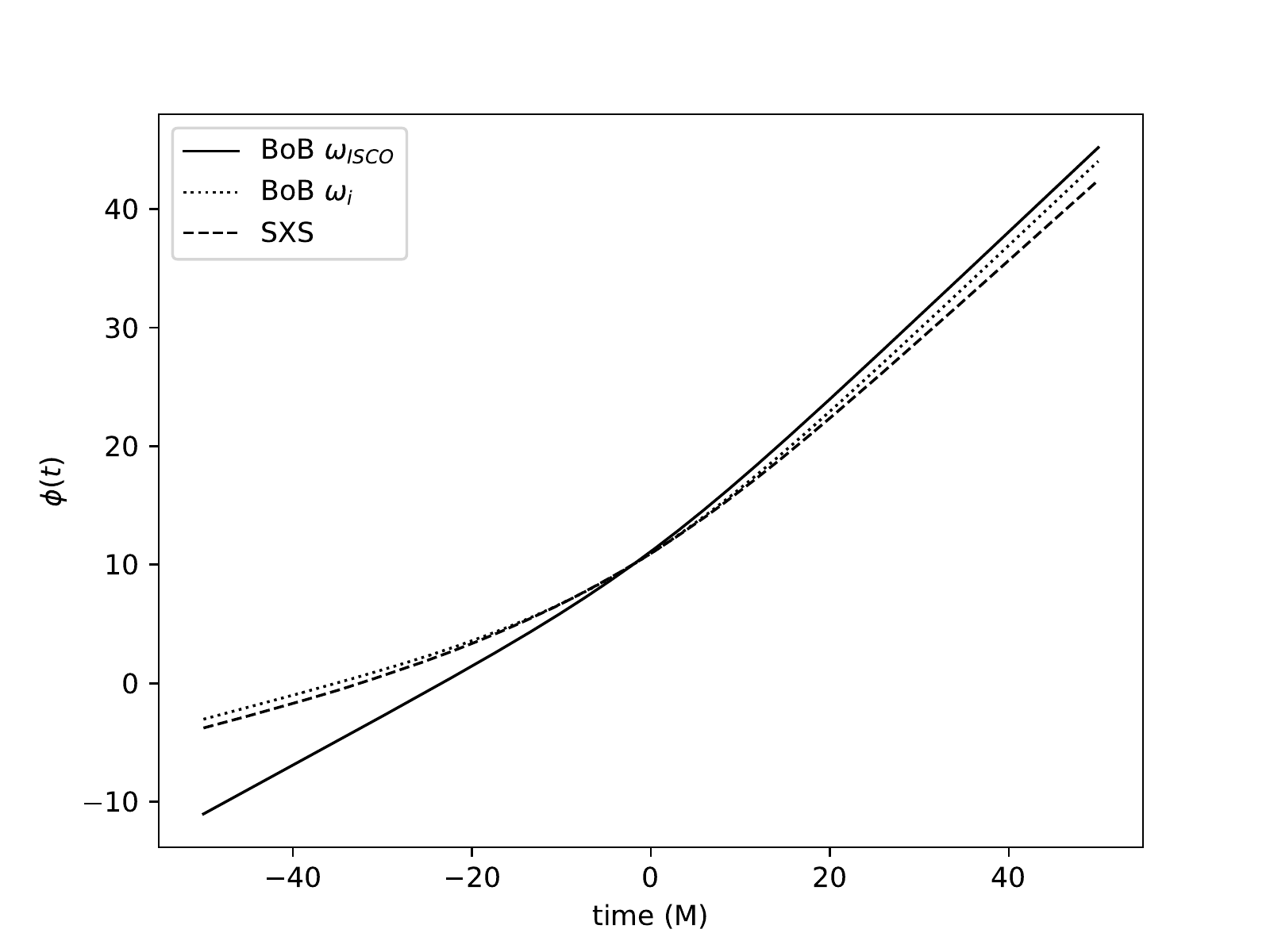}
\caption{The BoB phase for two initial frequencies compared with the SXS phase, aligned at the peak of frequency variation, for an equal-mass BBH with $\chi_f = 0.88$.}
\label{fig:phiBoBSXScomp}
\end{figure}

We compare next the analytic phase predicted by the BoB model with the two initial frequencies and the numerical phase from the SXS model, matched at the BoB peak of the increase in frequency $t_f$. 
We display the result in Fig.\ref{fig:phiBoBSXScomp}, which shows a reversed behavior: the BoB phase calculated with $\Omega_i$ gives a remarkable overall fit to numerical data. 
Even the initial difference in the phase computed with $\Omega_{ISCO}$ diminishes rapidly, such that during and after the merger it aligns well with the SXS phase, showing the robustness of the BoB model to variation in the initial data. 

\begin{figure}[!ht]
\centering
\includegraphics[scale=0.75]{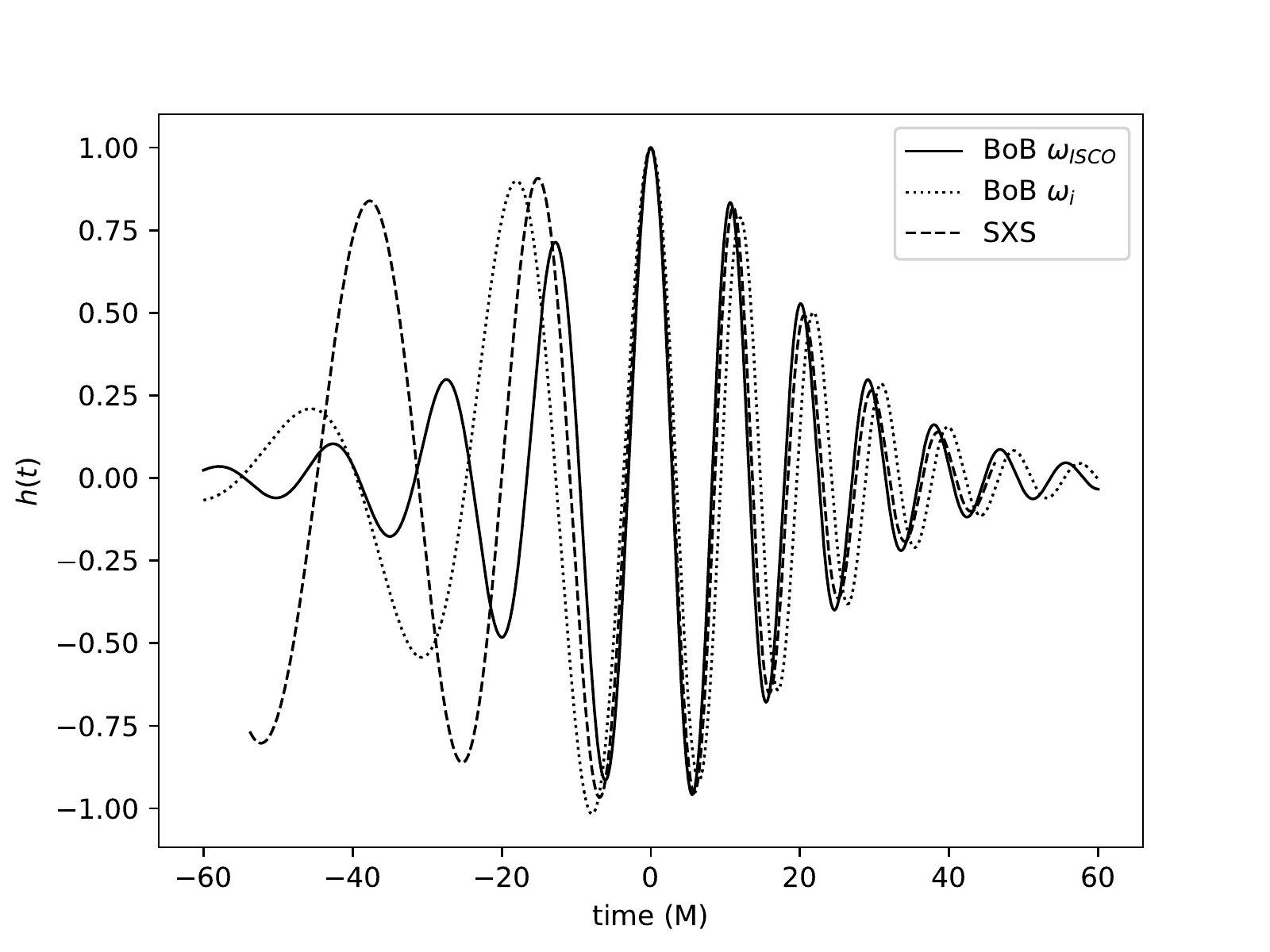}
\caption{The BoB frequency for two initial frequencies compared with the SXS frequency, aligned at the peak of frequency variation, for an equal-mass BBH with $\chi_f = 0.88$.}
\label{fig:omegaBoBSXScomp}
\end{figure}

We continue the comparison by taking the derivative of the SXS phase to obtain the evolution of the numerical frequency, filter it's inherent numerical noise, and plot it in Fig.\ref{fig:omegaBoBSXScomp} against the BoB frequency, for the two choices of initial frequency. We can see that the frequency calculated with $\Omega_{ISCO}$ gives a less accurate fit, and that in both situations BoB overestimates the ringdown frequency.

\begin{figure}[!ht]
\centering
\includegraphics[scale=0.75]{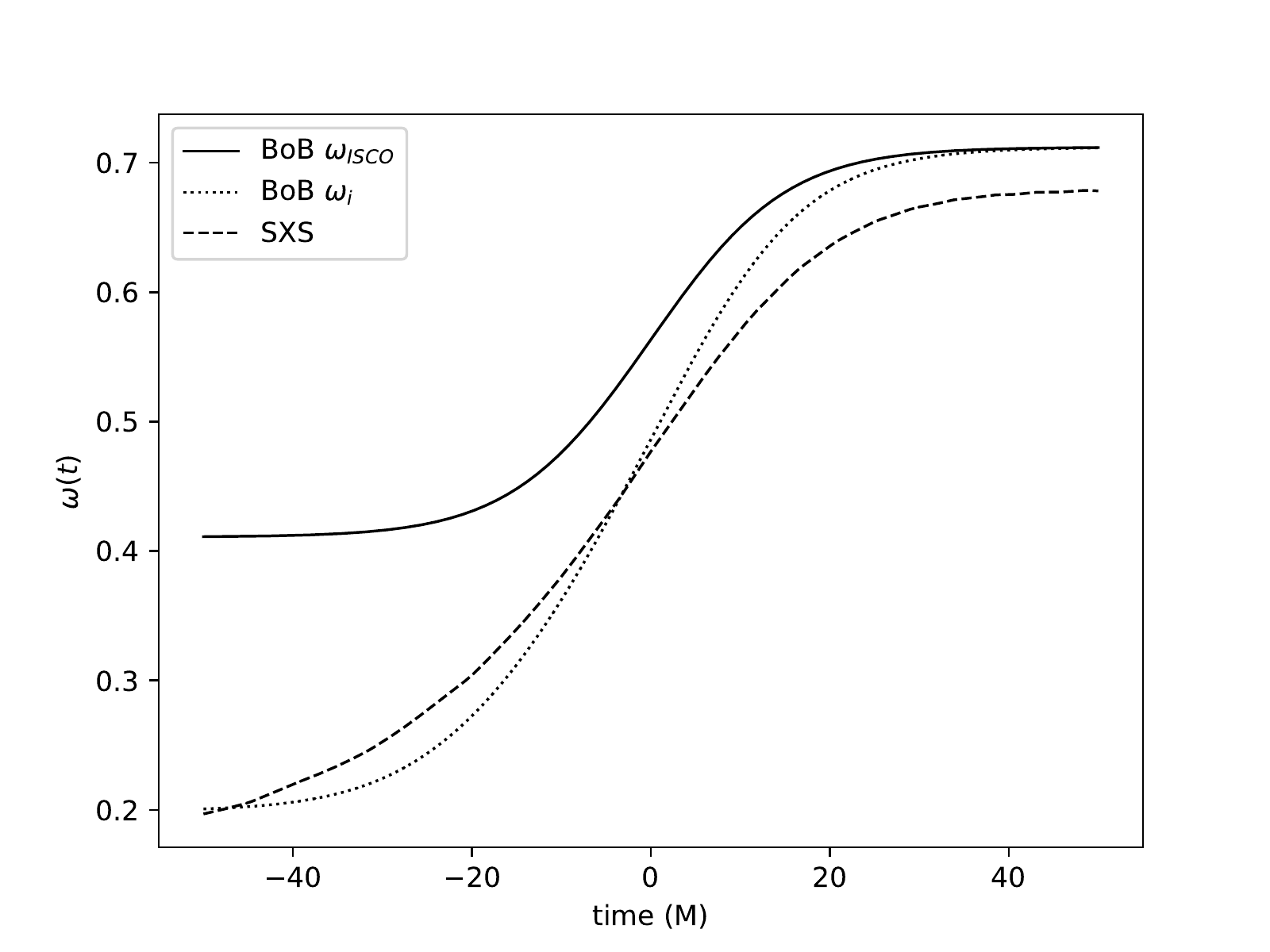}
\caption{Comparison between the BoB and SXS strains, matched at the highest peak, for an equal-mass binary merger with final spin $\chi_f = 0.88$.}
\label{fig:hBoBSXScomp}
\end{figure}

Finally, we build next the BoB strain and juxtapose it with the SXS template, matched at the highest peak in the strain amplitude and corrected with $\pi/2$ difference of phase, as shown in Fig.\ref{fig:hBoBSXScomp}. 
This phase difference we found between the numerical and analytical strain does not uncover a mismatch, because it is most likely due to a difference in the sign convention between the numerical and analytical calculations. 
We see in Fig.\ref{fig:hBoBSXScomp} that before the merger the mismatch is larger, due to the inaccuracy of the model to predict the signal before the LR, proved by the differences in amplitudes between the BoB and SXS waveforms that we uncovered. 
During the merger, the BoB model approximates well the SXS waveform for both choices of initial frequency. 
After the merger, both the amplitude and the period of the BoB strain start to deviate again from the numerical data, the BoB waveform calculated with $\omega_{ISCO}$ giving a better fit.

\section{Conclusion}
\label{sec:conclude}

In this work we implemented and tested two approaches that model the strong nonlinear gravitational interaction at the collision of two black holes, and computed analytically the GW signal.  Those models - the Implicit Rotating Source (IRS) and Backwards one Body (BoB) - are  related, the latter building up on the foundation laid by the former, by augmenting its heuristic approach with theoretical insights from the behavior of the GW signal around the LR.
Preliminary to building the merger waveform we presented and implemented an in-depth compendium of analytical methods available in the field for predicting the spin, mass, ring-down frequency and quality factor of the final black hole. 
Those quantities are essential because, as initial data, they influence the accuracy of the GW signal during the merger.

With the initial data known, we implemented and compared the IRS and BoB models for the merger with numerical data for non-spinning black holes, in the quasi-circular assumption.
We found an expected overall discrepancy in amplitude, an initial mismatch in frequency that gets resolved around the merger, and an excellent overall coincidence in phase, with a slight initial difference.
We obtained this agreement in phase after we accounted for a phase difference in the IRS waveform at the peak in the frequency variation.
With this correction, we achieved a remarkable agreement in the prediction of the GW strain for three different mass ratios between BoB and IRS.
We checked next the BoB model against the numerically calculated GW waveform for the spinning case, using our new implementation of the final spin, mass, frequency and quality factor, and performed a comparison between the analytically and numerically calculated GW. 
We mitigated for the uncertainty in choosing the initial frequency in the BoB model by employing two ways of calculating it and analyzed the consequences of our choices.
We found again a discrepancy in amplitude, more severe for our choice of initial frequency, a good agreement in phase, with a slight initial mismatch for the first, analytically estimated frequency, and a discrepancy in frequency, more pronounced for the original choice of initial frequency. 
We accounted for a $\pi/2$ phase difference to match the numerical and analytical strains at the peak and found a good accord between them, for both initial frequencies, from the LR onwards.
 
We conclude our findings by endorsing the BoB model because it depends only indirectly on the numerical data, through the analytical approximations used to estimate the mass, spin and quality factor of the final black hole. 
The greatest strength of this model is that it provides the initial time for the merger, which is invaluable when building complete analytical waveforms as we showed in \cite{arXiv:2203.08998}. 

Future research might include generic spin-precession, higher-order GW multipoles and recoil of the final black hole produced by the merger due to the asymmetric emission of GW, as well as thorough tests of the BoB model against the numerically-calculated waveforms, relying entirely on our implementation for the initial data, instead of using the numerically-supplied values for the final mass and spin.
The notebooks we developed in support of this work are available at \texttt{https://github.com/mbabiuc/MathScripts/tree/master/arXiv:2205.14742}.

\ack
We thank Dr. Nathan C. Steinle for useful suggestions. 
This research was supported in part by the National Science Foundation under EPSCoR Grant OIA-1458952 to the state of West Virginia “Waves of the Future” and Grant No. NSF PHY-1748958.

\section*{References}

\bibliography{references.bib}

\providecommand{\newblock}{}
\begin{thebibliography}{10}
\expandafter\ifx\csname url\endcsname\relax
  \def\url#1{{\tt #1}}\fi
\expandafter\ifx\csname urlprefix\endcsname\relax\def\urlprefix{URL }\fi
\providecommand{\href}[2]{#1}  
\providecommand{\eprint}[2][arXiv]{#1}

\bibitem{arXiv:1602.03837}
{Abbott} B~P {\em et~al\/} (LIGO Scientific, Virgo) 2016 {\em Phys. Rev.
  Lett.\/} {\bf 116} 061102 (\textit{Preprint}
  \eprint[http://arxiv.org/abs/1602.03837]{1602.03837})

\bibitem{arXiv:1608.01940}
{Abbott} B~P {\em et~al\/} 2017 {\em Annalen der Physik\/} {\bf 529} 1600209
  (\textit{Preprint} \eprint[http://arxiv.org/abs/1608.01940]{1608.01940})

\bibitem{arXiv:1411.4547}
{Aasi} J {\em et~al\/} (LIGO Scientific) 2015 {\em Classical and Quantum
  Gravity\/} {\bf 32} 074001 (\textit{Preprint}
  \eprint[http://arxiv.org/abs/1411.4547]{1411.4547})

\bibitem{arXiv:2111.03606}
{Abbott} B~P {\em et~al\/} (LIGO Scientific, Virgo, KAGRA) 2021 {\em arXiv
  e-prints\/}  arXiv:2111.03606 (\textit{Preprint}
  \eprint[http://arxiv.org/abs/2111.03606]{2111.03606})

\bibitem{arXiv:2112.06878}
{Nitz} A~H, {Kumar} S, {Wang} Y~F, {Kastha} S, {Wu} S, {Sch{\"a}fer} M,
  {Dhurkunde} R and {Capano} C~D 2021 {\em arXiv e-prints\/}  arXiv:2112.06878
  (\textit{Preprint} \eprint[http://arxiv.org/abs/2112.06878]{2112.06878})

\bibitem{arXiv:2010.14533}
Abbott R {\em et~al\/} (LIGO Scientific, Virgo) 2021 {\em Astrophys. J.
  Lett.\/} {\bf 913} L7 (\textit{Preprint}
  \eprint[http://arxiv.org/abs/2010.14533]{2010.14533})

\bibitem{arXiv:1811.12940}
Abbott B~P {\em et~al\/} (LIGO Scientific, Virgo) 2019 {\em Astrophys. J.
  Lett.\/} {\bf 882} L24 (\textit{Preprint}
  \eprint[http://arxiv.org/abs/1811.12940]{1811.12940})

\bibitem{book:exactgr}
Griffiths J~B and Podolsky J 2009 {\em {Exact Space-Times in Einstein's General
  Relativity}\/} Cambridge Monographs on Mathematical Physics (Cambridge:
  Cambridge University Press) ISBN 978-1-139-48116-8

\bibitem{arXiv:1411.3997}
Sperhake U 2015 {\em Class. Quant. Grav.\/} {\bf 32} 124011 (\textit{Preprint}
  \eprint[http://arxiv.org/abs/1411.3997]{1411.3997})

\bibitem{arXiv:2012.03608}
Hall E~D {\em et~al\/} 2021 {\em Phys. Rev. D\/} {\bf 103} 122004
  (\textit{Preprint} \eprint[http://arxiv.org/abs/2012.03608]{2012.03608})

\bibitem{arXiv:1912.02622}
Maggiore M {\em et~al\/} 2020 {\em JCAP\/} {\bf 03} 050 (\textit{Preprint}
  \eprint[http://arxiv.org/abs/1912.02622]{1912.02622})

\bibitem{arXiv:1907.11305}
Baker J {\em et~al\/} 2019 {Space Based Gravitational Wave Astronomy Beyond
  LISA} (\textit{Preprint}
  \eprint[http://arxiv.org/abs/1907.11305]{1907.11305})

\bibitem{arXiv:2001.09793}
Barausse E {\em et~al\/} 2020 {\em Gen. Rel. Grav.\/} {\bf 52} 81
  (\textit{Preprint} \eprint[http://arxiv.org/abs/2001.09793]{2001.09793})

\bibitem{arXiv:1802.06977}
Isoyama S, Nakano H and Nakamura T 2018 {\em PTEP\/} {\bf 2018} 073E01
  (\textit{Preprint} \eprint[http://arxiv.org/abs/1802.06977]{1802.06977})

\bibitem{arXiv:2008.10332}
Mei J {\em et~al\/} (TianQin) 2021 {\em PTEP\/} {\bf 2021} 05A107
  (\textit{Preprint} \eprint[http://arxiv.org/abs/2008.10332]{2008.10332})

\bibitem{arXiv:1907.04833}
Reitze D {\em et~al\/} 2019 {\em Bull. Am. Astron. Soc.\/} {\bf 51} 035
  (\textit{Preprint} \eprint[http://arxiv.org/abs/1907.04833]{1907.04833})

\bibitem{arXiv:2109.09882}
{Evans} M, {Adhikari} R~X, {Afle} C, {Ballmer} S~W, {Biscoveanu} S, {Borhanian}
  S, {Brown} D~A, {Chen} Y, {Eisenstein} R, {Gruson} A, {Gupta} A, {Hall} E~D,
  {Huxford} R, {Kamai} B, {Kashyap} R, {Kissel} J~S, {Kuns} K, {Landry} P,
  {Lenon} A, {Lovelace} G, {McCuller} L, {Ng} K~K~Y, {Nitz} A~H, {Read} J,
  {Sathyaprakash} B~S, {Shoemaker} D~H, {Slagmolen} B~J~J, {Smith} J~R,
  {Srivastava} V, {Sun} L, {Vitale} S and {Weiss} R 2021 {\em arXiv e-prints\/}
  arXiv:2109.09882 (\textit{Preprint}
  \eprint[http://arxiv.org/abs/2109.09882]{2109.09882})

\bibitem{arXiv:1607.05661}
Babak S, Taracchini A and Buonanno A 2017 {\em Phys. Rev. D\/} {\bf 95} 024010
  (\textit{Preprint} \eprint[http://arxiv.org/abs/1607.05661]{1607.05661})

\bibitem{arXiv:1611.00332}
Jim\'enez-Forteza X, Keitel D, Husa S, Hannam M, Khan S and P\"urrer M 2017
  {\em Phys. Rev. D\/} {\bf 95} 064024 (\textit{Preprint}
  \eprint[http://arxiv.org/abs/1611.00332]{1611.00332})

\bibitem{arXiv:2107.10193}
{Cristofoli} A, {Gonzo} R, {Kosower} D~A and {O'Connell} D 2021 {\em arXiv
  e-prints\/} arXiv:2107.10193 (\textit{Preprint}
  \eprint[http://arxiv.org/abs/2107.10193]{2107.10193})

\bibitem{arXiv:1808.06011}
{Duez} M~D and {Zlochower} Y 2019 {\em Reports on Progress in Physics\/} {\bf
  82} 016902 (\textit{Preprint}
  \eprint[http://arxiv.org/abs/1808.06011]{1808.06011})

\bibitem{arXiv:2101.11608}
{Tiglio} M and {Villanueva} A 2022 {\em Living Reviews in Relativity\/} {\bf
  25} 2 (\textit{Preprint}
  \eprint[http://arxiv.org/abs/2101.11608]{2101.11608})

\bibitem{arXiv:0805.1428}
Baker J~G, Boggs W~D, Centrella J, Kelly B~J, McWilliams S~T and van Meter J~R
  2008 {\em Phys. Rev. D\/} {\bf 78} 044046 (\textit{Preprint}
  \eprint[http://arxiv.org/abs/0805.1428]{0805.1428})

\bibitem{arXiv:1107.1181}
Kelly B~J, Baker J~G, Boggs W~D, McWilliams S~T and Centrella J 2011 {\em Phys.
  Rev. D\/} {\bf 84} 084009 (\textit{Preprint}
  \eprint[http://arxiv.org/abs/1107.1181]{1107.1181})

\bibitem{arXiv:1212.0837}
East W~E, McWilliams S~T, Levin J and Pretorius F 2013 {\em Phys. Rev. D\/}
  {\bf 87} 043004 (\textit{Preprint}
  \eprint[http://arxiv.org/abs/1212.0837]{1212.0837})

\bibitem{arXiv:1403.0561}
Huerta E~A, Kumar P, Gair J~R and McWilliams S~T 2014 {\em Phys. Rev. D\/} {\bf
  90} 024024 (\textit{Preprint}
  \eprint[http://arxiv.org/abs/1403.0561]{1403.0561})

\bibitem{arXiv:1403.7754}
Tai K~S, McWilliams S~T and Pretorius F 2014 {\em Phys. Rev. D\/} {\bf 90}
  103001 (\textit{Preprint} \eprint[http://arxiv.org/abs/1403.7754]{1403.7754})

\bibitem{arXiv:1810.00040}
McWilliams S~T 2019 {\em Phys. Rev. Lett.\/} {\bf 122}(19) 191102
  \urlprefix\url{https://link.aps.org/doi/10.1103/PhysRevLett.122.191102}

\bibitem{arXiv:2001.11412}
Pratten G, Husa S, Garcia-Quiros C, Colleoni M, Ramos-Buades A, Estelles H and
  Jaume R 2020 {\em Phys. Rev. D\/} {\bf 102} 064001 (\textit{Preprint}
  \eprint[http://arxiv.org/abs/2001.11412]{2001.11412})

\bibitem{arXiv:1304.6077}
Mroue A~H {\em et~al\/} 2013 {\em Phys. Rev. Lett.\/} {\bf 111} 241104
  (\textit{Preprint} \eprint[http://arxiv.org/abs/1304.6077]{1304.6077})

\bibitem{arXiv:1904.04831}
Boyle M {\em et~al\/} 2019 {\em Class. Quant. Grav.\/} {\bf 36} 195006
  (\textit{Preprint} \eprint[http://arxiv.org/abs/1904.04831]{1904.04831})

\bibitem{arXiv:1903.11704}
Reynolds C~S 2019 {\em Nature Astron.\/} {\bf 3} 41 (\textit{Preprint}
  \eprint[http://arxiv.org/abs/1903.11704]{1903.11704})

\bibitem{arXiv:2207.14290}
Gallegos-Garcia M, Fishbach M, Kalogera V, Berry C~P~L and Doctor Z 2022 {\em
  Astrophys. J. Lett.\/} {\bf 938} L19 (\textit{Preprint}
  \eprint[http://arxiv.org/abs/2207.14290]{2207.14290})

\bibitem{arXiv:2205.08574}
Callister T~A, Miller S~J, Chatziioannou K and Farr W~M 2022 {\em Astrophys. J.
  Lett.\/} {\bf 937} L13 (\textit{Preprint}
  \eprint[http://arxiv.org/abs/2205.08574]{2205.08574})

\bibitem{arXiv:2205.12329}
Mould M, Gerosa D, Broekgaarden F~S and Steinle N 2022 {\em Mon. Not. Roy.
  Astron. Soc.\/} {\bf 517} 2738 (\textit{Preprint}
  \eprint[http://arxiv.org/abs/2205.12329]{2205.12329})

\bibitem{arXiv:2010.00078}
Steinle N and Kesden M 2021 {\em Phys. Rev. D\/} {\bf 103} 063032
  (\textit{Preprint} \eprint[http://arxiv.org/abs/2010.00078]{2010.00078})

\bibitem{arXiv:2105.03439}
Gerosa D and Fishbach M 2021 {\em Nature Astron.\/} {\bf 5} 749
  (\textit{Preprint} \eprint[http://arxiv.org/abs/2105.03439]{2105.03439})

\bibitem{arXiv:gr-qc/9402014}
Cutler C and Flanagan E~E 1994 {\em Phys. Rev. D\/} {\bf 49} 2658
  (\textit{Preprint}
  \eprint[http://arxiv.org/abs/gr-qc/9402014]{gr-qc/9402014})

\bibitem{arXiv:gr-qc/9506022}
Kidder L~E 1995 {\em Phys. Rev. D\/} {\bf 52} 821 (\textit{Preprint}
  \eprint[http://arxiv.org/abs/gr-qc/9506022]{gr-qc/9506022})

\bibitem{arXiv:2103.03894}
Gangardt D, Steinle N, Kesden M, Gerosa D and Stoikos E 2021 {\em Phys. Rev.
  D\/} {\bf 103} 124026 (\textit{Preprint}
  \eprint[http://arxiv.org/abs/2103.03894]{2103.03894})

\bibitem{arXiv:1403.7377}
Will C~M 2014 {\em Living Rev. Rel.\/} {\bf 17} 4 (\textit{Preprint}
  \eprint[http://arxiv.org/abs/1403.7377]{1403.7377})

\bibitem{arXiv:1910.09908}
Ciufolini I, Paolozzi A, Pavlis E~C, Sindoni G, Ries J, Matzner R, Koenig R,
  Paris C, Gurzadyan V and Penrose R 2019 {\em Eur. Phys. J. C\/} {\bf 79} 872
  (\textit{Preprint} \eprint[http://arxiv.org/abs/1910.09908]{1910.09908})

\bibitem{arXiv:0904.2577}
Barausse E and Rezzolla L 2009 {\em Astrophys. J. Lett.\/} {\bf 704} L40
  (\textit{Preprint} \eprint[http://arxiv.org/abs/0904.2577]{0904.2577})

\bibitem{arXiv:1002.2643}
Kesden M, Sperhake U and Berti E 2010 {\em Phys. Rev. D\/} {\bf 81} 084054
  (\textit{Preprint} \eprint[http://arxiv.org/abs/1002.2643]{1002.2643})

\bibitem{arXiv:1312.5775}
Lousto C~O and Zlochower Y 2014 {\em Phys. Rev. D\/} {\bf 89} 104052
  (\textit{Preprint} \eprint[http://arxiv.org/abs/1312.5775]{1312.5775})

\bibitem{arXiv:1406.7295}
Healy J, Lousto C~O and Zlochower Y 2014 {\em Phys. Rev. D\/} {\bf 90} 104004
  (\textit{Preprint} \eprint[http://arxiv.org/abs/1406.7295]{1406.7295})

\bibitem{arXiv:1609.05933}
Huerta E~A {\em et~al\/} 2017 {\em Phys. Rev. D\/} {\bf 95} 024038
  (\textit{Preprint} \eprint[http://arxiv.org/abs/1609.05933]{1609.05933})

\bibitem{arXiv:1503.07536}
Zlochower Y and Lousto C~O 2015 {\em Phys. Rev. D\/} {\bf 92} 024022 [Erratum:
  Phys.Rev.D 94, 029901 (2016)] (\textit{Preprint}
  \eprint[http://arxiv.org/abs/1503.07536]{1503.07536})

\bibitem{arXiv:1610.09713}
Healy J and Lousto C~O 2017 {\em Phys. Rev. D\/} {\bf 95} 024037
  (\textit{Preprint} \eprint[http://arxiv.org/abs/1610.09713]{1610.09713})

\bibitem{arXiv:1612.02340}
Calder\'on~Bustillo J, Laguna P and Shoemaker D 2017 {\em Phys. Rev. D\/} {\bf
  95} 104038 (\textit{Preprint}
  \eprint[http://arxiv.org/abs/1612.02340]{1612.02340})

\bibitem{arXiv:1809.01401}
Tiwari V, Fairhurst S and Hannam M 2018 {\em Astrophys. J.\/} {\bf 868} 140
  (\textit{Preprint} \eprint[http://arxiv.org/abs/1809.01401]{1809.01401})

\bibitem{arXiv:1908.00555}
Fairhurst S, Green R, Hannam M and Hoy C 2020 {\em Phys. Rev. D\/} {\bf 102}
  041302 (\textit{Preprint}
  \eprint[http://arxiv.org/abs/1908.00555]{1908.00555})

\bibitem{arXiv:1909.05466}
Romero-Shaw I~M, Lasky P~D and Thrane E 2019 {\em Mon. Not. Roy. Astron.
  Soc.\/} {\bf 490} 5210 (\textit{Preprint}
  \eprint[http://arxiv.org/abs/1909.05466]{1909.05466})

\bibitem{arXiv:2010.04131}
Green R, Hoy C, Fairhurst S, Hannam M, Pannarale F and Thomas C 2021 {\em Phys.
  Rev. D\/} {\bf 103} 124023 (\textit{Preprint}
  \eprint[http://arxiv.org/abs/2010.04131]{2010.04131})

\bibitem{arXiv:1602.03840}
Abbott B~P {\em et~al\/} (LIGO Scientific, Virgo) 2016 {\em Phys. Rev. Lett.\/}
  {\bf 116} 241102 (\textit{Preprint}
  \eprint[http://arxiv.org/abs/1602.03840]{1602.03840})

\bibitem{arXiv:2009.01190}
Abbott R {\em et~al\/} (LIGO Scientific, Virgo) 2020 {\em Astrophys. J.
  Lett.\/} {\bf 900} L13 (\textit{Preprint}
  \eprint[http://arxiv.org/abs/2009.01190]{2009.01190})

\bibitem{arXiv:2009.01066}
Bustillo J~C, Sanchis-Gual N, Torres-Forn\'e A and Font J~A 2021 {\em Phys.
  Rev. Lett.\/} {\bf 126} 201101 (\textit{Preprint}
  \eprint[http://arxiv.org/abs/2009.01066]{2009.01066})

\bibitem{arXiv:2009.05461}
Gayathri V, Healy J, Lange J, O'Brien B, Szczepanczyk M, Bartos I, Campanelli
  M, Klimenko S, Lousto C and O'Shaughnessy R 2020 {\em arXiv e-prints\/}
  arXiv:2009.05461 (\textit{Preprint}
  \eprint[http://arxiv.org/abs/2009.05461]{2009.05461})

\bibitem{arXiv:2010.12558}
Nitz A~H and Capano C~D 2021 {\em Astrophys. J. Lett.\/} {\bf 907} L9
  (\textit{Preprint} \eprint[http://arxiv.org/abs/2010.12558]{2010.12558})

\bibitem{arXiv:1506.03492}
Gerosa D, Kesden M, Sperhake U, Berti E and O'Shaughnessy R 2015 {\em Phys.
  Rev. D\/} {\bf 92}(6) 064016 (\textit{Preprint}
  \eprint[http://arxiv.org/abs/1506.03492]{1506.03492})

\bibitem{arXiv:0710.3345}
Rezzolla L, Diener P, Dorband E~N, Pollney D, Reisswig C, Schnetter E and
  Seiler J 2008 {\em Astrophys. J. Lett.\/} {\bf 674} L29 (\textit{Preprint}
  \eprint[http://arxiv.org/abs/0710.3345]{0710.3345})

\bibitem{arXiv:0712.3541}
Rezzolla L, Barausse E, Dorband E~N, Pollney D, Reisswig C, Seiler J and Husa S
  2008 {\em Phys. Rev. D\/} {\bf 78} 044002 (\textit{Preprint}
  \eprint[http://arxiv.org/abs/0712.3541]{0712.3541})

\bibitem{arXiv:1508.07250}
Husa S, Khan S, Hannam M, P\"urrer M, Ohme F, Jim\'enez~Forteza X and Boh\'e A
  2016 {\em Phys. Rev. D\/} {\bf 93} 044006 (\textit{Preprint}
  \eprint[http://arxiv.org/abs/1508.07250]{1508.07250})

\bibitem{arXiv:1605.01938}
Hofmann F, Barausse E and Rezzolla L 2016 {\em Astrophys. J. Lett.\/} {\bf 825}
  L19 (\textit{Preprint} \eprint[http://arxiv.org/abs/1605.01938]{1605.01938})

\bibitem{arXiv:gr-qc/0103018}
Damour T 2001 {\em Phys. Rev. D\/} {\bf 64} 124013 (\textit{Preprint}
  \eprint[http://arxiv.org/abs/gr-qc/0103018]{gr-qc/0103018})

\bibitem{arXiv:0803.1820}
Racine E 2008 {\em Phys. Rev. D\/} {\bf 78} 044021 (\textit{Preprint}
  \eprint[http://arxiv.org/abs/0803.1820]{0803.1820})

\bibitem{arXiv:0706.3732}
Buonanno A, Pan Y, Baker J~G, Centrella J, Kelly B~J, McWilliams S~T and van
  Meter J~R 2007 {\em Phys. Rev. D\/} {\bf 76} 104049 (\textit{Preprint}
  \eprint[http://arxiv.org/abs/0706.3732]{0706.3732})

\bibitem{book:Wald}
Wald R~M 1984 {\em {General Relativity}\/} (Chicago, USA: Chicago Univ. Pr.)

\bibitem{christodoulou}
Christodoulou D 1970 {\em Phys. Rev. Lett.\/} {\bf 25}(22) 1596

\bibitem{smarr}
Smarr L 1973 {\em Phys. Rev. Lett.\/} {\bf 30} 71

\bibitem{arXiv:gr-qc/0206008}
Dreyer O, Krishnan B, Shoemaker D and Schnetter E 2003 {\em Physical Review
  D\/} {\bf 67} ISSN 1089-4918
  \urlprefix\url{http://dx.doi.org/10.1103/PhysRevD.67.024018}

\bibitem{arXiv:gr-qc/0306056}
Thornburg J 2003 {\em Classical and Quantum Gravity\/} {\bf 21} 743 ISSN
  1361-6382 \urlprefix\url{http://dx.doi.org/10.1088/0264-9381/21/2/026}

\bibitem{echeverria}
Echeverria F 1989 {\em Phys. Rev. D\/} {\bf 40} 3194

\bibitem{arXiv:gr-qc/0703053}
Berti E, Cardoso V, Gonzalez J~A, Sperhake U, Hannam M, Husa S and Br{\"u}gmann
  B 2007 {\em Physical Review D\/} {\bf 76} ISSN 1550-2368
  \urlprefix\url{http://dx.doi.org/10.1103/PhysRevD.76.064034}

\bibitem{arXiv:0907.0462}
Reisswig C, Husa S, Rezzolla L, Dorband E~N, Pollney D and Seiler J 2009 {\em
  Phys. Rev. D\/} {\bf 80} 124026 (\textit{Preprint}
  \eprint[http://arxiv.org/abs/0907.0462]{0907.0462})

\bibitem{arXiv:gr-qc/9909058}
Kokkotas K~D and Schmidt B~G 1999 {\em Living Reviews in Relativity\/} {\bf 2}
  ISSN 1433-8351 \urlprefix\url{http://dx.doi.org/10.12942/lrr-1999-2}

\bibitem{arXiv:1404.3197}
London L, Shoemaker D and Healy J 2014 {\em Phys. Rev. D\/} {\bf 90} 124032
  [Erratum: Phys.Rev.D 94, 069902 (2016)] (\textit{Preprint}
  \eprint[http://arxiv.org/abs/1404.3197]{1404.3197})

\bibitem{arXiv:0905.2975}
Berti E, Cardoso V and Starinets A~O 2009 {\em Class. Quant. Grav.\/} {\bf 26}
  163001 (\textit{Preprint} \eprint[http://arxiv.org/abs/0905.2975]{0905.2975})

\bibitem{arXiv:1408.1860}
Berti E and Klein A 2014 {\em Phys. Rev. D\/} {\bf 90} 064012
  (\textit{Preprint} \eprint[http://arxiv.org/abs/1408.1860]{1408.1860})

\bibitem{arXiv:gr-qc/0512160}
Berti E, Cardoso V and Will C~M 2006 {\em Physical Review D\/} {\bf 73} ISSN
  1550-2368 \urlprefix\url{http://dx.doi.org/10.1103/PhysRevD.73.064030}

\bibitem{arXiv:0903.0338}
{Sathyaprakash} B~S and {Schutz} B~F 2009 {\em Living Reviews in Relativity\/}
  {\bf 12} 2 (\textit{Preprint}
  \eprint[http://arxiv.org/abs/0903.0338]{0903.0338})

\bibitem{arXiv:2001.10914}
Garc\'\i{}a-Quir\'os C, Colleoni M, Husa S, Estell\'es H, Pratten G,
  Ramos-Buades A, Mateu-Lucena M and Jaume R 2020 {\em Phys. Rev. D\/} {\bf
  102} 064002 (\textit{Preprint}
  \eprint[http://arxiv.org/abs/2001.10914]{2001.10914})

\bibitem{mashhoon}
Ferrari V and Mashhoon B 1984 {\em Phys. Rev. D\/} {\bf 30} 295

\bibitem{BertiWeb}
Berti E Ringdown \urlprefix\url{https://pages.jh.edu/eberti2/ringdown}

\bibitem{newman}
Newman E and Penrose R 1962 {\em J. Math. Phys.\/} {\bf 3} 566

\bibitem{arXiv:1006.1632}
Reisswig C and Pollney D 2011 {\em Class. Quant. Grav.\/} {\bf 28} 195015
  (\textit{Preprint} \eprint[http://arxiv.org/abs/1006.1632]{1006.1632})

\bibitem{arXiv:1711.06276}
Huerta E~A {\em et~al\/} 2018 {\em Phys. Rev. D\/} {\bf 97} 024031
  (\textit{Preprint} \eprint[http://arxiv.org/abs/1711.06276]{1711.06276})

\bibitem{book:optics}
Born M, Wolf E, Bhatia A~B, Clemmow P~C, Gabor D, Stokes A~R, Taylor A~M,
  Wayman P~A and Wilcock W~L 1999 {\em Principles of Optics: Electromagnetic
  Theory of Propagation, Interference and Diffraction of Light\/} 7th ed
  (Cambridge University Press)

\bibitem{arXiv:1703.00118}
An X 2017 {\em arXiv e-prints\/} arXiv:1703.00118 (\textit{Preprint}
  \eprint[http://arxiv.org/abs/1703.00118]{1703.00118})

\bibitem{arXiv:2202.08848}
Gerosa Davide~Fabbri C~M and Sperhake U 2022 {\em arXiv e-prints\/}
  arXiv:2202.08848 (\textit{Preprint}
  \eprint[http://arxiv.org/abs/2202.08848]{2202.08848})

\bibitem{bardeen}
Bardeen J~M, Press W~H and Teukolsky S~A 1972 {\em Astrophys. J.\/} {\bf 178}
  347

\bibitem{arXiv:gr-qc/0001013}
Buonanno A and Damour T 2000 {\em Physical Review D\/} {\bf 62} ISSN 1089-4918
  \urlprefix\url{http://dx.doi.org/10.1103/PhysRevD.62.064015}

\bibitem{arXiv:2203.08998}
{Buskirk} D and {Babiuc-Hamilton} M 2022 {\em arXiv e-prints\/}
  arXiv:2203.08998 (\textit{Preprint}
  \eprint[http://arxiv.org/abs/2203.08998]{2203.08998})

\bibitem{arXiv:1912.11716}
Abbott R {\em et~al\/} (LIGO Scientific, Virgo) 2021 {\em SoftwareX\/} {\bf 13}
  100658 (\textit{Preprint}
  \eprint[http://arxiv.org/abs/1912.11716]{1912.11716})

\bibitem{arXiv:0904.3541}
Lousto C~O, Campanelli M, Zlochower Y and Nakano H 2010 {\em Class. Quant.
  Grav.\/} {\bf 27} 114006 (\textit{Preprint}
  \eprint[http://arxiv.org/abs/0904.3541]{0904.3541})

\bibitem{arXiv:1604.00778}
Nielsen A~B 2016 {\em J. Phys. Conf. Ser.\/} {\bf 716} 012002
  (\textit{Preprint} \eprint[http://arxiv.org/abs/1604.00778]{1604.00778})

\bibitem{arXiv:1611.03703}
Boh\'e A {\em et~al\/} 2017 {\em Phys. Rev. D\/} {\bf 95} 044028
  (\textit{Preprint} \eprint[http://arxiv.org/abs/1611.03703]{1611.03703})

\bibitem{arXiv:1612.09566}
Keitel D {\em et~al\/} 2017 {\em Phys. Rev. D\/} {\bf 96} 024006
  (\textit{Preprint} \eprint[http://arxiv.org/abs/1612.09566]{1612.09566})

\bibitem{arXiv:1708.00404}
London L, Khan S, Fauchon-Jones E, Garc\'\i{}a C, Hannam M, Husa S,
  Jim\'enez-Forteza X, Kalaghatgi C, Ohme F and Pannarale F 2018 {\em Phys.
  Rev. Lett.\/} {\bf 120} 161102 (\textit{Preprint}
  \eprint[http://arxiv.org/abs/1708.00404]{1708.00404})

\end{thebibliography}

\clearpage

\begin{appendix}

\section{Formulas and Coefficients for the Final Spin}
\label{appendixA}

A phenomenological formulation for the spin of the final black hole is proposed in 
\cite{arXiv:0710.3345, arXiv:0907.0462} and revised in \cite{arXiv:1605.01938}.
Another empirical formula for the spin, mass and recoil of the final black hole in terms of mass ratios and individual spins is developed in \cite{arXiv:0904.3541, arXiv:1312.5775} based on an analytical fit to numerical simulations data using Taylor expansions, and is extended in \cite{arXiv:1406.7295, arXiv:1503.07536, arXiv:1610.09713}. 
We implement these approaches in eqs.\eref{eq:chifin1} and \eref{eq:chifin2}.

A third analytical formulation, based on a hierarchical approach, is introduced in \cite{arXiv:1508.07250, arXiv:1611.00332}. 
They start from the dimensionless orbital angular momentum at ISCO, $L_{ISCO}$.
Although this quantity does depend implicitly on $\chi_f$, they find a workaround by rewriting it as a polynomial in $\eta$ and $\chi_{eff}$ with coefficients informed from numerical data. Then, the final spin is recovered with the formula $\chi_f = {\tilde L_{ISCO}} + \chi_1 + \chi_2$. 
We work through this hierarchical approach presented in \cite{arXiv:1508.07250, arXiv:1611.00332} and give in eq.\eref{eq:chifin3} the lengthly expression we used. 

The first formula used for the spin is from \cite{arXiv:0710.3345}.
\begin{eqnarray}
\label{eq:chifin1}
&\chi_{f,1}=s_{00}+\eta s_{01}+\eta^2 s_{02}+\eta^3 s_{03}+s_{10} \chi_{eff} \nonumber\\
&+\eta s_{11} \chi_{eff}+\eta^2 s_{12} \chi_{eff}+s_{20} \chi_{eff}^2+\eta s_{21}\chi_{eff}^2+s_{30}\chi_{eff}^3
\end{eqnarray}
The coefficients entering in the equation \eref{eq:chifin1} are given below, in eq\eref{eq:ascoefs}: 
\begin{eqnarray}
\label{eq:ascoefs}
&s_{00} = s_{20}=s_{30}=0,~s_{10}=1,~s_{01} = 2\sqrt{3},~s_{21} =-0.1229,\nonumber\\
&s_{12}=0.4537,~s_{11}=-2.8904,~s_{02}= -3.5171,~s_{03}= 2.5763.
\end{eqnarray}
The second, alternative formula used in calculating the spin are from \cite{arXiv:1406.7295, arXiv:1610.09713} 
\begin{eqnarray}
\label{eq:chifin2}
&\chi_{f2}=16 \mu ^2 (\delta_{\chi}^4 L_{4c}+\delta_{\chi}^3 \delta_m L_{4b}+\delta_{\chi}^2 L_{2c}+\delta_{\chi}^2 L_{3b} \sigma_{\chi}+\delta_{\chi}^2 L_{4e} \sigma_{\chi}^2+\delta_{\chi}^2 \delta_m^2 L_{4h}\nonumber \\
&+\delta_{\chi} \delta_m L_{2a} +\delta_{\chi} \delta_m L_{3a} \sigma_{\chi}+\delta_{\chi} \delta_m L_{4a} \sigma_{\chi}^2+\delta_{\chi} \delta_m^3 L_{4g}+L_{0}+L1 \sigma_{\chi} +L_{2b} \sigma_{\chi}^2+\delta_m^2 L_{2d} \nonumber \\
&+L_{3c} \sigma_{\chi}^3+\delta_m^2 L_{3d} \sigma_{\chi}+L_{4d} \sigma_{\chi}^4+\delta_m^4 L_{4f}+\delta_m^2 L_{4i} \sigma_{\chi}^2 )+\delta_m^4 (8\mu +1) \sigma_{\chi}.
\end{eqnarray}
Here, 
\begin{equation}
\delta_m=\frac{m_1-m_2}{m},~
\delta_{\chi}=\frac{\chi_2-\chi_1 q}{1+q},~
\sigma_{\chi}=\frac{\chi_2+\chi_1 q^2}{(1+q)^2}
\end{equation}
We use two sets of fits for the coefficients entering in the equation (A.3), the first one given in Table VI of \cite{arXiv:1406.7295} and the second from Table III of \cite{arXiv:1610.09713}.

The third formula for the spin is taken from \cite{arXiv:1611.00332}
\begin{eqnarray}
\label{eq:chifin3}
&\chi_{f2}=\frac{\eta  (a_{10}+\eta  (a_{2} a_{20}+a_{3} a_{30} \eta))}{a_{5} a_{50} \eta +1} \nonumber \\
&+\bigg (\eta \chi_{eff} (b_{1} b_{10} (-4 \eta^2 (4 f_{11}+f_{12}-16)+f_{11}+f_{12} \eta )  \nonumber \\
&+b_{2} b_{20} \chi_{eff} (-4 \eta ^2 (4 f_{21}+f_{22}-16)+f_{21}+f_{22} \eta ) \nonumber \\
&+b_{3} b_{30}\chi_{eff}^2(-4 \eta ^2 (4 f_{31}+f_{32}-16)+f_{31}+f_{32} \eta )) \bigg) \nonumber \\
&\bigg / \bigg ( b_{5} b_{50} \chi_{eff} (-4 \eta^3 (16 f_{50}+4 f_{51}+f_{52}-16)+f_{50}+f_{51} \eta +f_{52} \eta^2)+1\bigg ) \nonumber \\
&+\Delta d_{10} \eta^2 \chi_{diff} (d_{11} \eta +1)+d_{20} \eta^3 \chi_{diff}^2\nonumber \\
&+\Delta d_{30} \eta^3 \chi_{diff} \chi_{eff} (d_{31} \eta +1)+\chi_{tot}.
\end{eqnarray}
Again, there are two fits to numerical data results for the coefficients, presented below:
\begin{eqnarray}
\label{eq:Hcoefs1}
&a_{10} = 2 \sqrt{3},~ a_{20} = 5.28,~a_{30} = 1.27,~a_{50} = 2.89,\nonumber \\
&b_{10} =-0.194,~b_{20} = 0.075,~b_{30} = 0.00782,~b_{50} = -0.527,\nonumber \\
&a_{2} =3.7724,~a_{3} = -9.6278 ,~a_{5} = 2.4874,\nonumber \\
&b_{1} = 1.0005 ,~b_{2} = 0.88234 ,~b_{3} = 0.76128 ,~b_{5} = 0.91392 ,\nonumber \\
&d_{10} = 0.27628 ,~d_{11} = 11.562,~d_{20} = -0.059758 ,~d_{30} = 2.7297 ,~d_{31} = -3.3883 ,\nonumber \\
&f_{11} = 4.4110 ,~f_{12}  =  0.36422 , ~f_{21}  =  8.8879 ,~f_{22}  =  -40.354 , \nonumber \\
&f_{31}  = 23.927 ,~f_{32}  =  -178.78 , ~f_{50}  =  1.8982 , ~f_{51}  =  -5.5570 , ~f_{52}  =  0
\end{eqnarray}
\begin{eqnarray}
\label{eq:Hcoefs2}
&a_{10} = 2 \sqrt{3},~ a_{20} = 5.24,~a_{30} = 1.3,~a_{50} = 2.88,\nonumber \\
&b_{10} =-0.194,~b_{20} = 0.0851,~b_{30} = 0.00954,~b_{50} = -0.579,\nonumber \\
&a_{2} =3.8326,~a_{3} = -9.4874,~a_{5} = 2.5135,\nonumber \\
&b_{1} = 1.00096 ,~b_{2} = 0.78775 ,~b_{3} = 0.65401,~b_{5} = 0.83967 ,\nonumber \\
&d_{10} = 0.32237 ,~d_{11} = 9.3326,~d_{20} = -0.059808 ,~d_{30} = 2.31704 ,~d_{31} = -3.2625 ,\nonumber \\
&f_{11} = 4.4092 ,~f_{12}  =  0.51183 , ~f_{21}  =  8.7737 ,~f_{22}  =  -32.061 , \nonumber \\
&f_{31}  = 22.830 ,~f_{32}  =  -153.84 , ~f_{50}  =  1.88047, ~f_{51}  =  -4.7702 , ~f_{52}  =  0
\end{eqnarray}

To test our implementation we choose as a first example an equal-mass binary with a total mass normalized to $1M$, in which each black hole has the same dimensionless spin $\chi_1=\chi_2 = 0.5$, situated on the lower-end of stellar-mass black holes spins \cite{arXiv:1604.00778}. 
From eq.\eref{eq:chieff} the effective spin will be $\chi_{eff} = 0.5$, and using the three methods mentioned above we obtain $\chi_f = 0.8314\pm 0.00014$.

\section{Formulas and Coefficients for the Final Mass}
\label{appendixB}

An empirical relation for the final mass was introduced in \cite{arXiv:0904.3541}, refined in \cite{arXiv:1312.5775, arXiv:1406.7295} to include binaries with different masses and spins, and recalibrated in \cite{arXiv:1503.07536, arXiv:1610.09713}. 
A different phenomenological model for the analytical fit to the radiated energy is built in \cite{arXiv:1611.00332}, based on an hierarchical approach starting from the radiated energy of a test-particle that plunges into a rotating (or Kerr) black hole after passing ISCO, which is known analytically \cite{bardeen, arXiv:gr-qc/0001013}.
\begin{equation}
{\tilde E_{ISCO}}= \eta \frac{{\tilde r_{ISCO}^{3/2}} - 2{\tilde r^{1/2}_{ISCO}} + \chi_f}
{{\tilde r_{ISCO}^{3/4}}\sqrt{{\tilde r_{ISCO}^{3/2}} - 3{\tilde r^{1/2}_{ISCO}} + 2 \chi_f}}
\label{eq:EISCO}
\end{equation} 
This model creates a three-dimensional parameter space for the distribution of mass and spin, which is calibrated with both NR and PN results \cite{arXiv:1611.03703, arXiv:1612.09566} and is subsequently refined to include spin precession in \cite{arXiv:1708.00404, arXiv:2001.11412}.     
We implement these fits for the energy of the GW and compare their predictions for the final black hole mass. 

From \cite{arXiv:0907.0462}, we pick the formula for estimating the radiated energy at the collision of an equal-mass spin-aligned binary black holes system, namely eq.\eref{eq:MfAEI}, with two sets of coefficients given in eq.\eref{eq:MfAEIcoefs}, and eq.\eref{eq:MfRIT} with coefficients given in eq.\eref{eq:MfRITcoefs}.
\begin{equation}
\label{eq:MfAEI}
M^{eq}_{f}= M (1 - \tilde E_{GW}) = M\left ( 1 - \sum_{i=0}^2 p_{i} (\chi_1 + \chi_2)^i \right)
\end{equation}
where 
\begin{equation}
\label{eq:MfAEIcoefs}
p_0 = \frac{4.826}{100},~ p_1 = \frac{1.559}{100},~ p_2 = \frac{0.485}{100},
\end{equation}
and
\begin{equation}
\label{eq:MfRIT}
M^{eq}_{f}= M\left (1 - q_0 - q_1 (\chi_1 + \chi_2) - q_2 (\chi_1 + \chi_2)^2- q_3 (\chi_1 - \chi_2)^2\right)
\end{equation}
where 
\begin{equation}
\label{eq:MfRITcoefs}
q_0 = \frac{5.025}{100},~ q_1 = \frac{1.352}{100},~ q_2 =- \frac{0.0219}{100}, q_3 = -\frac{0.270}{100}.
\end{equation}
From \cite{arXiv:1312.5775} we gather an improved fit for the radiated energy, valid also only for equal-mass binaries with aligned spins, written in eq.\eref{eq:MfRIT1}:
\begin{equation}
\label{eq:MfRIT1}
M^{eq}_{f}= M\left(1  - E_0 - E_2 \chi_f^2 - E4 \chi_f^4 \right)
\end{equation}
with coefficients $(E_1, E_2, E_3)$ given in Table III of \cite{arXiv:1312.5775}.
One more formula for the radiated energy in GW is given in \eref{eq:MfRIT2}, with the first sets of coefficients listed in Table VI of \cite{arXiv:1406.7295} and the second in Table III of \cite{arXiv:1610.09713}.
\begin{eqnarray}
\label{eq:MfRIT2}
&M_f  = 16 \mu^2 \bigg ( \delta_{\chi}^4 K_{4c} + \delta_{\chi}^3 \delta_m K_{4b} + \delta_{\chi}^2 K_{2c} + \delta_{\chi}^2 K_{3b} \sigma_{\chi} +\delta_{\chi}^2 K_{4e} \sigma_{\chi}^2 + \delta_{\chi}^2 \delta_m^2 K_{4h} \nonumber \\
&  + \delta_{\chi} \delta_m K_{2a} + \delta_{\chi} \delta_m K_{3a} \sigma_{\chi}+\delta_{\chi} \delta_m K_{4a} \sigma_{\chi}^2 + \delta_{\chi} \delta_m^3 K_{4g} + K_{1} \sigma +K_{2b} \sigma^2 + \delta_m^2 K_{2d} \nonumber \\
& + K_{3c} \sigma_{\chi}^3 + \delta_m^2 K_{3d} \sigma_{\chi} + K_{4d} \sigma_{\chi}^4 + \delta_m^4 K_{4f} + \delta_m^2 K_{4i} \sigma_{\chi}^2 + M_0 \bigg )
\end{eqnarray}

Lastly, we implement a fifth formula for the mass, taken from \cite{arXiv:1611.00332}
\begin{eqnarray}
\label{eq:MfH}
&M_f  =  M(1 -  d_{10} \Delta \eta^2 \chi_{diff} (1+d_{11} \eta) - d_{20} \eta^3 \chi_{diff}^2 - d_{30} \Delta \eta \chi_{diff} \chi_{eff} (1+ d_{31} \eta)  \nonumber \\
& + \eta (a_1 + \eta (a_2 + \eta (a_3 + a_4 \eta)) (1+ 0.346 b_1 \chi_{eff} (f_{10} + \eta (f_{11} +  f_{12} \eta))  \nonumber \\
&+  0.211 b_2 \chi_{eff}^2 (f_{20} + \eta (f_{21} + f_{22} \eta)) + 0.128 b_3 \chi_{eff}^3 (f_{30} + \eta (f_{31} + f_{32} \eta)))) \nonumber \\
& \bigg / (1 - 0.212 b_5 \chi_{eff} (f_{50} + \eta (f_{51} + f_{52} \eta)))  )
\end{eqnarray}
with the coefficients:
\begin{eqnarray}
\label{eq:MfHcoefs}
&a_1 = 0.057191, ~a_2 = 0.5610, ~a_3 = -0.847,~ a_4 = 3.145, \nonumber \\
&b_1 = -0.209,~ b_2 = -0.197, ~b_3 = -0.159, ~b5 = 2.985, \nonumber \\
&d_{10} =-0.098, ~d_{11} = -3.23, ~d_{20} = 0.0112,~d30 =-0.0198, ~d_{31} = -4.92,\nonumber \\
&f_{10} = 1.80908,~ f_{11} = 15.7,~ f_{12} = -75.74528, ~f_{20} = 4.27, \nonumber \\
&f_{21} = 0,~f_{22} = -52.448, ~f_{30} = 31.09, ~f_{31} =-243.6,\nonumber \\
&f_{32} = 492.96, ~f_{50} = 1.56735,~ f_{51} = -0.58, ~f_{52} = -6.7576.
\end{eqnarray}
We test all these formulas for the simplest case of an equal-mass non-spinning system and obtain 
$M_f =0.9516 \pm 0.00015$.
For unequal masses we can use only  eqs.\eref{eq:MfRIT2} and \eref{eq:MfH}. 

 
\end{appendix}

\end{document}